\documentclass[reprint,amsmath,amssymb,aapm,onecolumn]{revtex4-2}
\usepackage{graphicx}
\usepackage{dcolumn}
\usepackage{bm}
\usepackage{booktabs}
\usepackage{tabularray}
\usepackage{float} 
\begin{document}

\title {Pseudo grid-based physics-informed convolutional-recurrent network solving the integrable nonlinear lattice equations}                     
\vspace{3.8in}
\author{Zhe Lin}
\affiliation{School of Mathematical Sciences, Key Laboratory of Mathematics and Engineering Applications (Ministry of Education) and Shanghai Key Laboratory of PMMP, East China Normal University, Shanghai 200241, China} 
\author{Yong Chen}
\email{ychen@sei.ecnu.edu.cn}
\affiliation{School of Mathematical Sciences, Key Laboratory of Mathematics and Engineering Applications (Ministry of Education) and Shanghai Key Laboratory of PMMP, East China Normal University, Shanghai 200241, China} 

\affiliation{College of Mathematics and Systems Science, Shandong University of Science and Technology, Qingdao 266590, China}

\vspace{10.8in}
\begin{abstract}
\baselineskip=11pt
\noindent{\it Abstract:}
Traditional discrete learning methods involve discretizing continuous equations using difference schemes, necessitating considerations of stability and convergence. Integrable nonlinear lattice equations possess a profound mathematical structure that enables them to revert to continuous integrable equations in the continuous limit, particularly retaining integrable properties such as conservation laws, Hamiltonian structure, and multiple soliton solutions. The pseudo grid-based physics-informed convolutional-recurrent network (PG-PhyCRNet) is proposed to investigate the localized wave solutions of integrable lattice equations, which significantly enhances the model's extrapolation capability to lattice points beyond the temporal domain. We conduct a comparative analysis of PG-PhyCRNet with and without pseudo grid by investigating the multi-soliton solutions and rational solitons of the Toda lattice and self-dual network equation. The results indicate that the PG-PhyCRNet excels in capturing long-term evolution and enhances the model's extrapolation capability for solitons, particularly those with steep waveforms and high wave speeds. Finally, the robustness of the PG-PhyCRNet method and its effect on the prediction of solutions in different scenarios are confirmed through repeated experiments involving pseudo grid partitioning.
\\
\\
\noindent{\it Keywords: Pseudo grid-based physics-informed convolutional-recurrent network, Integrable nonlinear lattice equations, Multi-soliton solutions, Rational soliton solutions} 
\end{abstract}

\maketitle
\baselineskip=13pt

\section{Introduction}
Solving systems of continuous and discrete nonlinear partial differential equations (PDEs) has been a key component in scientific computing and mathematical physics. In recent years, deep learning has achieved great success in solving PDEs due to the powerful representation capabilities of neural networks~\cite{pinn,FNO,deeponet}. Research on modeling mathematical-physical systems using deep neural networks can be broadly classified into two categories: continuous networks and discrete networks. Both continuous and discrete integrable systems, a special class of nonlinear systems, possess a rich mathematical structure and unique properties, such as Lax pairs, Hamiltonian structures, and infinite conservation laws, distinguishing them from ordinary nonlinear systems~\cite{integrable,NLE1,NLE2}.

\subsection{Continuous and discrete learning methods}
Significant progress has been made in the study of nonlinear system surrogate modeling using deep learning~\cite{NN1,disc,FNO,deeponet}. Many studies focus on minimizing the physical information error function, driving the vibrant growth of scientific machine learning research, particularly in the modeling and simulation of PDEs~\cite{pinn,pideeponet}. Karniadakis' group proposed a landmark data-driven approach based on fully connected neural networks (FCNs): Physics-Informed Neural Networks (PINNs), which is an efficient and powerful method for solving the forward and inverse problems of PDEs\cite{pinn}. Compared to purely data-driven neural network learning, PINN imposes physical information constraints during training, allowing for the learning of more generalizable models with fewer data samples and the wide application in aerodynamic flows, solid mechanics, blood flows modeling and many others~\cite{apinn1,apinn2,apinn3,apinn4}. Chen's group proposed the concept and framework of integrable deep learning and conducted a series of significant studies. Many data-driven localized wave solutions of continuous integrable equations can be obtained by PINN~\cite{integrable_pinn1,integrable_pinn2,integrable_pinn3,integrable_pinn33,integrable_pinn44,integrable_pinn55}. Besides, they investigated some complex problems in integrable systems using the improved PINN method: high-dimensional integrable systems, nonlocal integrable equations, the forward and inverse problem  for variable coefficients integrable equations, and solving the higher-order RW solutions~\cite{integrable_pinn4,integrable_pinn5,integrable_pinn6,integrable_pinn11}. There are also some innovative works that integrate the perfect mathematical structures of integrable systems. For example, the two-stage PINN based on conservation laws can enhance the prediction accuracy and generalization ability of solutions~\cite{integrable_pinn7}, the PINN based on the Miura transformation discovered a new kink-bell solution of modified Korteweg de-Vries (mKdV) equation~\cite{integrable_pinn8}, and the LPNN based on Lax pairs accelerated training efficiency and improves the prediction accuracy of localized wave solutions~\cite{integrable_pinn10}. Additionally, some other excellent papers on integrable deep learning that readers can refer to Refs.~\cite{integrable_pinn12,integrable_pinn13,integrable_pinn14,integrable_pinn15}. 

Unlike FCNs, convolutional neural networks (CNNs) extract spatial hierarchical features through local connections and weight sharing, significantly reducing the number of parameters and computational complexity. This makes CNNs more efficient for handling large-scale data and high-dimensional problems, and particularly well-suited for surrogate modeling of discretized PDEs. Discretized learning approaches based on CNNs have demonstrated good convergence and scalability in modeling PDEs~\cite{UQ,phygeonet,dense,breakingdata,self_conv,physr,phycnn,phycrnet}. Researchers applied CNNs to surrogate modeling and uncertainty quantification of PDEs in rectangular reference domains~\cite{UQ}. PhyGeoNet achieved geometrically adaptive solutions for steady-state partial differential equations through coordinate transformations~\cite{phygeonet}. The AR-DenseED method successfully solved PDEs using discretized learning without labeled data~\cite{dense}. The physics-informed convolutional-recurrent network (PhyCRNet) method, based on an encoder-decoder convolutional long short-term memory (LSTM) network, inherited the advantages of AR-DenseED in feature extraction in low-dimensional space and time evolution learning, demonstrating superior performance in the accuracy, extrapolation, and generalization of PDEs' solutions~\cite{phycrnet}. 

\subsection{Discrete integrable systems and deep learning}
The pioneering study on nonlinear lattice equations (NLEs) can be traced back to the early 1950s, when Fermi, Pasta, and Ulam (FPU) conducted great numerical experiments~\cite{FPU1}. They established a model of vibrating particles connected by nonlinear springs and, through computer simulations, discovered that waves with various sinusoidal initial conditions would return to their initial energy states after a certain period. To extract quantitative information from models described by continuous differential equations, it is often necessary to solve them numerically using various discretization methods. However, it is generally expected that the discretization of any given integrable equations may lead to a non-integrable system. It is crucial to assure that the discretized models exhibit the same qualitative features of the dynamics as their continuous counterparts. Therefore, Toda investigated the motion equations of a one-dimensional uniform lattice with nearest-neighbor interactions and propose the integrable Toda lattice~\cite{NLE1,NLE2,Toda1,Toda2,Toda3}. The existence of analytical solutions in integrable lattice equations opens up new avenues for a better understanding of nonlinear phenomena, not only in the context of FPU recurrence but also in other areas related to nonlinear mechanics~\cite{AL1,ZZN,ZDJ}. Until the 1970s, the topic of integrable discretization continued to advance. Case and Kac~\cite{transd1} focused on the discretization of the Schrödinger spectral problem and the discretization of the Ablowitz-Kaup-Newell-Segur spectral problem (developed by Ablowitz and Ladik~\cite{NLE2,AL1}) as well as establishing a difference-based inverse scattering transformation. Hirota~\cite{transd2,transd3} utilized the bilinear Bäcklund transformations and their connection with Lax pairs to derive a series of discrete integrable systems for bilinear equations. Subsequent research on NLEs have expanded, including the nonlinear self-dual network equation under nonlinear inductor-capacitor (LC) circuits~\cite{selfd1,selfd2}, Ablowitz-Ladik (AL) lattice~\cite{AL1,NLE2}, Błaszak-Marciniak lattice~\cite{BM}, Volterra lattice~\cite{VL}, etc. These equations have made significant theoretical contributions to describing physical phenomena such as crystalline phenomena and nonlinear lattice dynamics\cite{app3,app5}. Methods for solving these lattice equations analytically, such as the inverse scattering transformation method~\cite{AL1}, Bäcklund transformation~\cite{bkl}, Hirota method~\cite{hirota1,hirota2}, and Darboux transformation~\cite{DT2}, have been developed to better understand these mathematical structures of lattice solutions and physical phenomena.

Discrete integrable systems share many similarities with their continuous counterparts, which possess a pair of linear spectral problems (known as discrete Lax pairs), infinite conservation law, Hamiltonian structure, symmetries, and other features. In this paper, some significant discrete integrable systems can converge to their corresponding continuous integrable systems by taking the appropriate continuum limit:
\begin{itemize}
  \item The first equation discussed in this paper is the Toda lattice~\cite{NLE1}:
\begin{equation}
\frac{\mathrm{d}^2}{\mathrm{d}t^2}\left(\ln u_n\right)=u_{n+1}+u_{n-1}-2u_n,
\end{equation}
or another coupled form:
\begin{equation}\label{intro_toda}
  \dot{\bf{p}}_n = \left(\begin{array}{c}
                     v_n-v_{n-1} \\
                     u_{n+1}-u_{n} 
                   \end{array}\right),
\end{equation}
where ${\bf p}_n=(\ln u_n, v_n)^T$, $\dot{{\bf p}}_n=\mathrm{d}{\bf p}_n/\mathrm{d}t$. Inspired by Refs.~\cite{transd2,transd3,lxjx}, we give a step-size parameter $\delta$ and transformation
\begin{equation}\label{toda_cl}
  u_n(t)=1+\delta q\left(\delta n -6\delta t, \delta^2t\right):=1+\delta q\left(x, \tau\right), \ v_n(t)=1+\delta p\left(\delta n +\frac{\delta^2}{4} t, \delta t\right):=1+\delta p\left(y, k\right).
\end{equation}
Eliminating the linear terms in the system of equations, the lattice~(\ref{intro_toda}) can approximate: 
\begin{equation}\label{kdv}
  (q_{\tau}+q_{xxx}+6qq_x)\delta^3+\mathcal{O}(\delta^4)=0,
\end{equation}
which is just the famous Korteweg de-Vries (KdV) equation when the higher-order terms are neglected. Eq.~(\ref{kdv}) was first derived by Korteweg and de-Vries in the study of long water waves in relatively shallow channels, which has important applications in nonlinear science and mathematical physics~\cite{kdv1,kdv2}.
  \item The second equation discussed is self-dual network eqaution~\cite{hirota_sd}:
\begin{equation}\label{intro_sd}
\dot{\bf{y}}_n = \left[\begin{array}{c}
                     (1+I_n^2)(V_{n-1}-V_n) \\
                     (1+V_n^2)(I_{n}-I_{n+1}) 
                   \end{array}\right],
\end{equation}
where ${\bf y}_n=(I_n, V_n)^T$. We give the transformation
\begin{equation}\label{sd_cl}
I_n(t)=\alpha+\Delta(\alpha)q(x,\tau), \ V_n(t)=\alpha+\Delta(\alpha)p(x,\tau),
\end{equation}
with 
\begin{equation*}
  \Delta(\alpha)=\frac{(\alpha^2+1)\delta}{\sqrt{1-3\alpha^2}}, \ x=\delta n +b\delta t, \ \tau=-\frac{1}{6}b\delta^3t, \ 0<\alpha<1, b\in\mathbb{R},
\end{equation*}
the lattice~(\ref{intro_sd}) can approximate: 
\begin{equation}\label{mkdv}
  (q_{\tau}+q_{xxx}+6q^2q_x)\delta^4+\mathcal{O}(\delta^5)=0,
\end{equation}
which is just the mKdV when the higher-order terms are neglected. The mKdV equation was first introduced by Miura and can serve as a model equation for acoustic waves in anharmonic lattices~\cite{miura}. It can also describe dust-acoustic solitary waves in dusty plasmas and wave phenomena in nonlinear optics~\cite{mkdv1,mkdv2}.
\end{itemize}
The step-size parameter $\delta$ measures the grid size of the spatial component and allows us to bring lattice points closer together to create a continuous point. It is not difficult to observe that higher-order dispersion terms in continuous integrable equations transform into a simple translation relationship under integrable discretization. This greatly facilitates our numerical computations. From this perspective, we are considering whether discrete learning methods in deep learning can be used to further investigate the dynamical behavior of solutions in the integrable lattice equations.

To the best of our knowledge, researches on using deep learning to solve problems related to NLEs are relatively limited. Saqlain et al. have addressed the inverse problem of parameter identification in discrete high-dimensional systems by appropriately adjusting the PINN method~\cite{inverse_discrete,AL_PINN}. In data-driven mainstream learning processes, interpolation and extrapolation are involved in evaluating model performance. Interpolation requires the generalisation ability within the boundaries of the training data. Extrapolation refers to estimating unknown values by extending beyond the known range based on known sequences or factors~\cite{in_ex}. PINN learns the function mappings directly in spatio-temporal regions, which leads to a lack of spatiotemporal extrapolation ability due to not focusing on the temporal evolution of physical processes. However, the discrete learning method of PhyCRNet discretizes PDEs and establishes dependencies between temporal data through CNNs and LSTM, ensuring that the model solution possesses a certain degree of extrapolation ability in the temporal direction. Besides, CNNs focus on the discrete points of the input tensor, making them more suitable for exploring discrete systems. Compared to continuous learning methods like PINN, we prefer to use discrete learning methods that directly target the discrete grid points in lattice equations to explore the dynamical behavior of the solutions. Enhancing extrapolation is often achieved by altering the model structure or providing more prior conditions. The pseudo grid-based physics-informed convolutional-recurrent network (PG-PhyCRNet) method, which incorporates pseudo grid training, is based on the richer prior conditions (initial value conditions) in the theory of discrete integrable systems. It addresses the issue of irregular distribution of integer lattice, which causes the network to fail to capture the potential function features and leads to prediction failures. Thus, the motivation and highlights of this paper are:
\begin{itemize}
  \item Our motivation is to utilize PhyCRNet methods to directly explore integrable NLEs, which is a novel attempt. In the traditional PhyCRNet learning methods, it is essential to carefully consider the stability and convergence of the discretization of continuity equations using finite difference schemes (especially for higher-order dispersion and nonlinear terms). However, as a unique discrete scheme of its integrable continuous equations, NLEs possess a richer mathematical structure and retains the same integrable properties. Moreover, under the continuous limit~(\ref{toda_cl}) and (\ref{sd_cl}), this may allow an indirect discussion of the dynamical behavior of continuous solutions with the help of discrete solutions.
  \item PG-PhyCRNET significantly enhances the model's extrapolation capability to lattice points beyond the temporal domain. Based on more prior information about initial values obtained by the integrable nature of discrete lattice equations, PG-PhyCRNet enables easier to learn the dynamical features of lattice solitons.
  \item The PG-PhyCRNet approach offers superior performance in capturing the dynamics of high-speed, steep-profile solitons, particularly in extrapolating solutions beyond the initial time domain. Multi-soliton and rational soliton solutions of the Toda lattice and self-dual network equations are investigated by using the PhyCRNet method with and without pseudo grid (PG-PhyCRNet and PhyCRNet). The performance of these two training methods in different scenarios are comparatively analysed: the PhyCRNet method is better suited for lattice solitons with lower speeds and smoother profiles; the PG-PhyCRNet method is better suited for lattice solitons with higher speeds and steeper profiles, where training with a pseudo grid gives an accuracy improvement in the prediction of the lattice solution outside the time domain. 
\end{itemize}

The structure of this paper is as follows: Section~\ref{sec2} provides a model description and  details PG-PhyCRNet method. Sections~\ref{sec3} and \ref{sec4} discuss the numerical experiments of the PhyCRNet method without and (PG-PhyCRNet and PhyCRNet) in predicting different types of lattice solitons, respectively. In Section~\ref{sec5}, the effect of pseudo grid in PG-PhyCRNet method and the stability of these localized waves are analyzed through repeated experiments. Finally, the last section provides a discussion and summary of this paper.

\section{Model description}\label{sec2}
In this section, we will introduce the PG-PhyCRNet for solving $(1+1)$-dimensional semi-discrete integrable systems. The architecture's ability to handle different solutions of semi-discrete integrable systems from two perspectives are discussed. Both perspectives involve extracting low-dimensional spatial features and learning the temporal evolution of compressed information through an encoder-decoder convolutional-recurrent scheme. 

\subsection{1D-ConvLSTM}
ConvLSTM is an architecture that combines CNNs and LSTMs specifically designed for processing sequential data~\cite{lstm,convlstm}. Unlike traditional LSTMs, ConvLSTM applies convolution operations at each time step, which helps capture spatial information in the sequential data~\cite{convlstm}. It consists mainly of input gates, forget gates, output gates, and convolutional layers. In the gating mechanism, ConvLSTM uses convolutional layers to compute the relevance between the input at the current time step and the hidden state from the previous time step, which is used to control the input and output information at the current time step and to forget the information from the previous hidden state. 

\begin{figure*}[!htb]
    \begin{center}
    {\scalebox{0.7}[0.7]{\includegraphics{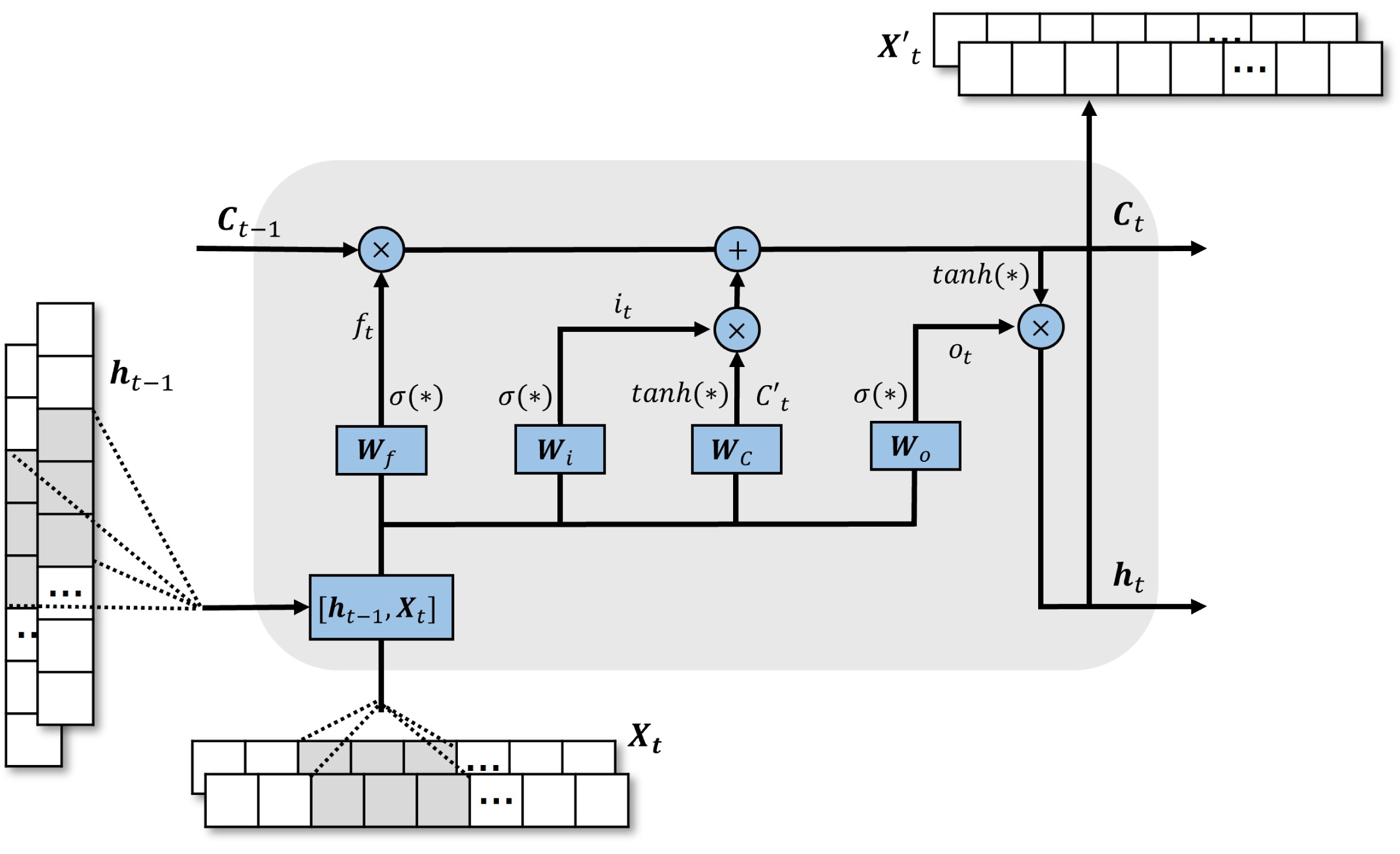}}}
    \end{center}
    \vspace{-0.15in} \caption{\small Chain-like structure of ConvLSTM.} \label{convlstm}
\end{figure*}

ConvLSTM also implements three gates (the forget gate ${\bf f}_t$, input gate ${\bf i}_t$, and output gate ${\bf o}_t$) to protect and control the cell state. Let ${\bf X}_t$ be the input tensor, and $\{{\bf h}_{t-1}, {\bf C}_t\}$ be the hidden state and cell state to be updated at time step $t$, as shown in the chain-like structure in Fig.~\ref{convlstm}. The forgetting gate ${\bf f}_t$ reads the information from $\{{\bf h}_{t-1}, {\bf X}_t\}$ and passes through the sigmoid layer $\sigma(\ast)$ to obtain a value between $0$ and $1$. When the forget gate is close to $1$, the information in the cell state is fully retained; otherwise, the information is completely forgotten. This adaptive behavior allows the forget gate to clear memorized information in the cell state ${\bf C}_{t-1}$. The input gate ${\bf i}_t$ has two parts: the first sigmoid layer decides what to update, and the second tanh layer updates the information into the cell state. The output gate ${\bf o}_t$ uses a sigmoid layer determines how much of the current cell state to output. The mathematical formula for the ConvLSTM unit can be expressed as:
\begin{equation}\label{lstm}
\begin{aligned}
  {\bf i}_t &= \sigma({\bf W_i}\ast[{\bf X}_t, {\bf h}_{t-1}]+{\bf b}_i), \ \ {\bf f}_t = \sigma({\bf W_f}\ast[{\bf X}_t, {\bf h}_{t-1}]+{\bf b}_f),\\
  \widetilde{{\bf C}}_{t-1} &= \tanh({\bf W_c}\ast[{\bf X}_t, {\bf h}_{t-1}]+{\bf b}_c), \ \ {\bf C}_t = {\bf f}_t\odot{\bf C}_{t-1}+{\bf i}_{t}\odot\widetilde{{\bf C}}_{t-1},\\
  {\bf o}_t &= \sigma({\bf W_o}\ast[{\bf X}_t, {\bf h}_{t-1}]+{\bf o}_i), \ \ {\bf h}_t = {\bf o}_t\odot\tanh({\bf C}_t),
\end{aligned}
\end{equation}
where $\ast$ is the convolutional operation and $\odot$ denotes the Hadamard product. The presence of ConvLSTM ensures that in the process of solving integrable lattice equations, the dependence relationship between the potential function and the preceding and succeeding states is preserved. Moreover, the information flowing into ConvLSTM is compressed and latent features are extracted by an encoder.

\subsection{Network Architecture}
\begin{figure*}[!htb]
    \begin{center}
    {\scalebox{0.23}[0.23]{\includegraphics{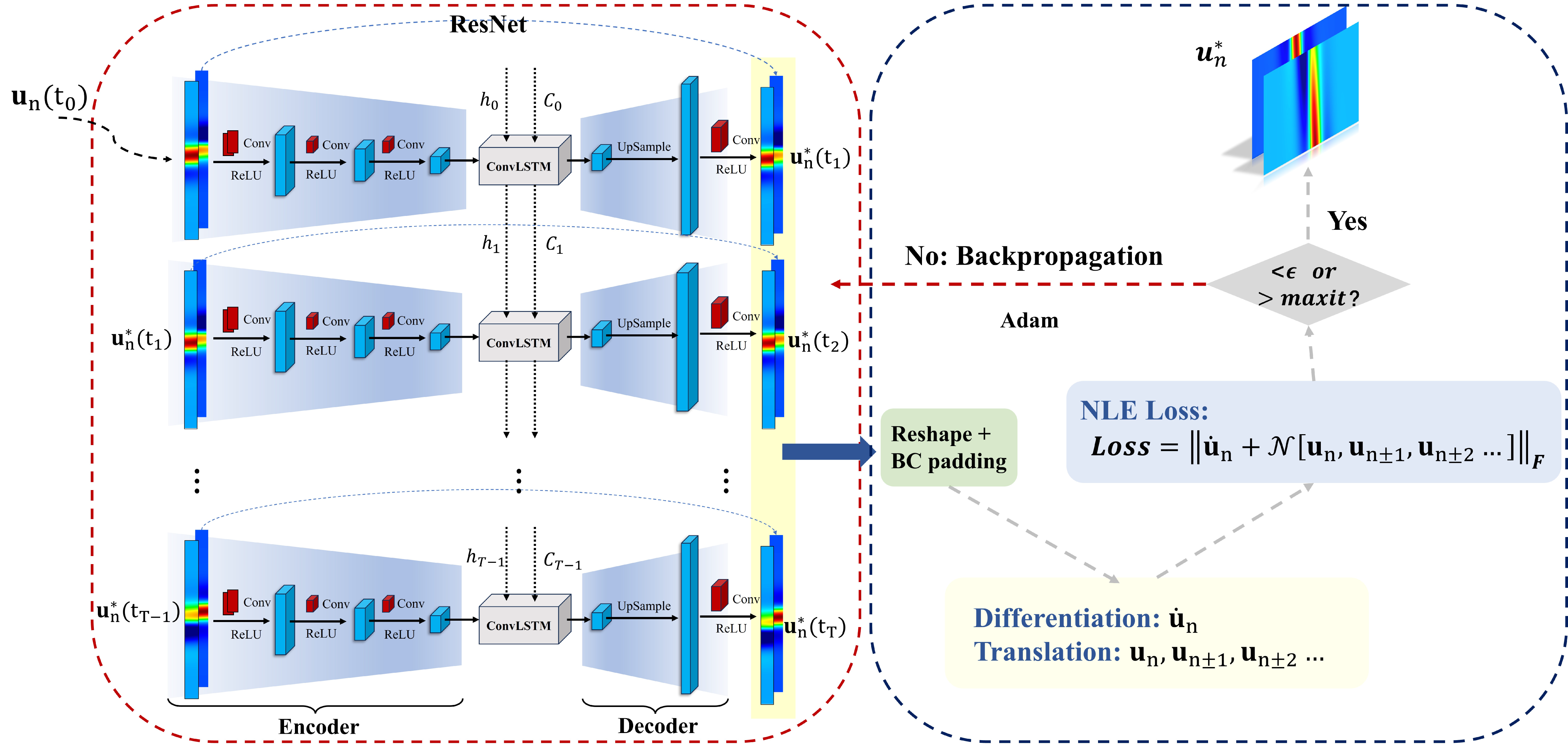}}}
    \end{center}
    \vspace{-0.15in} \caption{\small Network architecture. "Reshape" represents the reshaping of the potential function tensor at each epoch. "BC padding" represents the boundary condition. } \label{flow}
\end{figure*}

PhyCRNet is a discrete learning method for solving multi-dimensional spatiotemporal PDEs without any labeled data~\cite{phycrnet}. The architecture mainly includes an encoder-decoder structure, residual connections, autoregressive processes, and differentiation based on gradient-free convolutional kernels. The exploration of integer lattice points in integrable NLEs is essential. Therefore, the initial consideration is to directly apply PhyCRNet to train and predict integer lattice points. However, due to the irregular distribution of many lattice solitons near the wave peaks, more prior information needs to be incorporated into the model for joint training. This approach can lead to better predictions for integer lattice points. Therefore, we propose the PG-PhyCRNet method to solve the integrable lattice equation in two different situation:
\begin{itemize}
  \item We prefer to focus on the dynamical behavior of integer lattice sites in the spatial part of the NLE. Hence, we illustrate this approach using the initial–boundary value problem of (1+1)-dimensional NLE as an example:
\begin{equation}\label{nle}
    \begin{cases}
        \dot{u}_n + \mathcal{N}[u_n, u_{n\pm1}, u_{n\pm2}...]=0,\\
        u_{n}(0)=u(n,0), \ n\in\Omega,\\
        u_{n_1}= u_{b1}, \ u_{n_N}= u_{b2}, \ t\in[t_0, t_T],
    \end{cases}
\end{equation}
where $u_n=u(n,t)$ denotes the solution of NLE, $\mathcal{N}$ is the nonlinear operator, $n\in\mathbb{Z}$ stands for lattice site and the region $\Omega=[n_1, n_N]$ is a finite one-dimension lattice, $u_{b_1}$ and $u_{b_2}$ are constants.

  \item If the additional information beyond the initial values of the integer lattice sites can be obtained, such as the initial value samples of the integer lattice sites following a certain distribution or empirical regression function (i.e., the initial condition in Eq.~(\ref{nle}) are changed to $u_{n}(0)=u(n,0), \ n\in\Omega', \Omega'=[n_1, n_1+i\frac{n_N-n_1}{N_n-1}],\ i=1,...,N_n-1$), then the trend of integer lattice sites with help with non-integer grid can be studied. We refer to these non-integer grid points as "pseudo grid" of the network. This operation is actually to better fit the continuous potential function $u(n,t)$ with respect to variables $n$ and $t$ through model training. For cases like solitary waves with fast wave speeds and narrow wave widths, it is often difficult to capture their propagation characteristics near the peak using integer lattice sites. The addition of pseudo grid avoids this problem and enables the solution $u_n$ based on the well-trained model to have good extrapolation ability in the time direction. However, the drawback of this method is that it may lose prediction accuracy within the given region. 
\end{itemize}
By utilizing the exact solutions of discrete integrable systems to provide rich training samples, both training methods under different prior conditions can be satisfied. Fig.~\ref{flow} illustrates the PG-PhyCRNet architecture for solving discrete integrable systems.
The encoder structure consists of three convolutional operations, which are used to learn low-dimensional latent features from the input state variables $u_{n}(t_i) (i = 0,1,...)$, with a ReLU function applied after each convolution for non-linearity. The ConvLSTM layer receives the feature information passed by the encoder and propagates the temporal feature information through the latent features $h_i$ and cell states $C_i$. The decoder includes a pixel shuffle operation \cite{pix} and a convolution operation with a large kernel size, aiming to scale the low-resolution feature information passed by the ConvLSTM layer back to the original size of interest and reconstruct it as the information for the next time step in high resolution. Finally, a simple autoregressive process is achieved by using residual connections to update the potential function states for adjacent time steps, where $u_{n}(t_{i+1})=u_{n}(t_{i})+\delta_t*\mathcal{N}\mathcal{N}[\cdot]$, where $\mathcal{N}\mathcal{N}$ denotes the trained network operator and $\delta_t$ is the time interval.     

\subsection{Translation operation and derivative approximation }
After obtaining the output from the model on the left side of Fig.~\ref{flow}, the outputs are concatenated at each time step and the boundary conditions (BCs) is hard imposed. This is done to facilitate the subsequent use of translation operation and gradient-free convolution filters to represent the translation operations ($E^kf_n=f_{n+k}, n, k\in \mathbb{Z}$) in the spatial direction and discrete numerical differentiation in the time direction. 

\begin{figure*}[!htb]
    \begin{center}
    {\scalebox{0.5}[0.5]{\includegraphics{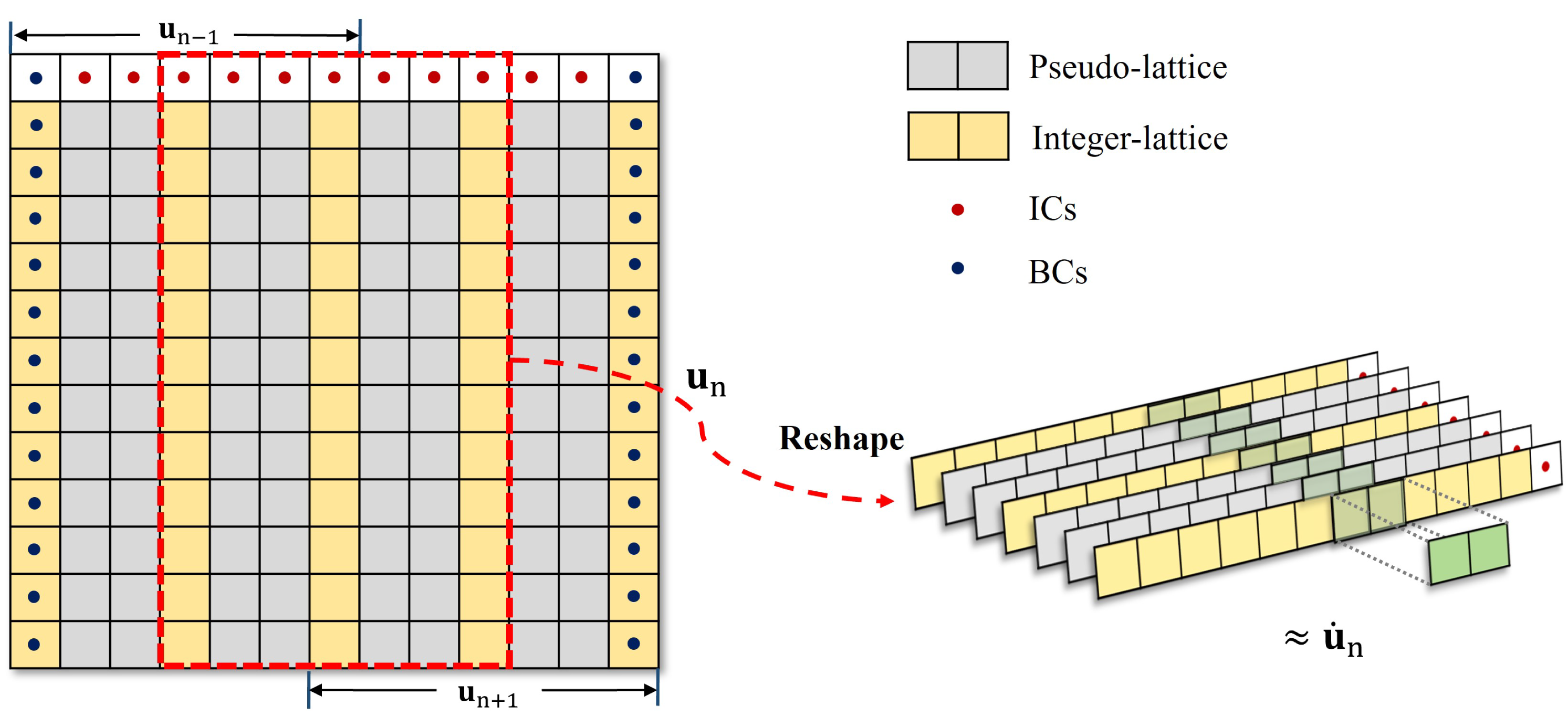}}}
    \end{center}
    \vspace{-0.15in} \caption{\small The translation operations in the spatial direction (left) and the discrete numerical differentiation in the time direction (right).} \label{diff}
\end{figure*}

Due to the higher-order dispersion terms in continuous integrable equations can transform into a simple translation relationship under integrable discretization, the translation operator is very important as a surrogate in the network. We give the following two ways of handling the translation operator based on model training methods without and with pseudo grid:\\
\noindent(1) Translated potential function $u_{n\pm1}, u_{n\pm2}, \ldots$ can be obtained by directly shifting the concatenated predicted values. It is worth noting that for the first training method without adding pseudo grid, we simply shift the predicted solution forward or backward by $k$ steps to obtain $u_{n\pm k}$. \\
\noindent(2) For the training method with pseudo grid, we need to shift the predicted solution forward or backward by $k/\delta_n \ (\delta_n\in\mathcal{A})$ steps to obtain $u_{n\pm k}$, where 
\begin{equation}
  \mathcal{A}=\left\{\delta_n\left| \delta_n=\frac{n_N-n_1}{N_n-1} ,N_n\geq2, N_n\in\mathbb{Z_+}, \frac{1}{\delta_n}\in\mathbb{Z_+}\right.\right\}.
\end{equation}
The corresponding processes of tensor recombination, hard-imposed of BCs, and translation operators are illustrated in Fig.~\ref{diff}. The left side in Fig.~\ref{diff} shows the translation operation in the training method with pseudo grid, where the red dashed part represents the potential function $u_n$ that will enter the loss function. And the translation operator in NLEs can be replaced by a proper slicing operation on the output.

Besides, we use forward differential format from the finite difference method to approximate the derivatives in the time direction. From the right side of Fig.~\ref{diff}, it can be seen that we decompose the two-dimensional potential function into a one-dimensional potential function that varies with time. We approximate the derivative in the time direction by applying a one-dimensional gradient-free convolution kernel of size $2$ to the reorganized potential function.

\subsection{Parameter optimization and capability evaluation of PG-PhyCRNet}
The essence of training a neural network or deep learning model lies in updating its weights and biases. Based on the training data, our goal is to minimize the value of the loss function by optimizing the parameters of the neural network. The presence of the loss function indicates the direction of training for deep learning models. 

Due to the imposition of strict initial boundary conditions, there is no longer a need to include these conditions as constraints in the loss function. Therefore, the loss function only needs to define the required physical information to be satisfied. Building upon the two training methods mentioned in the previous section, we uniformly define their residual $\mathcal{R}({\bf n}, t; {\bf \theta})$ as:
\begin{equation}\label{residual}
  \mathcal{R}({\bf n}, t; {\bf \theta})=\dot{{\bf u}}_n + \mathcal{N}[{\bf u}_n, {\bf u}_{n\pm1}, {\bf u}_{n\pm2}...].
\end{equation}
This expression actually represents the left-hand side of Eq.~(\ref{nle}), where ${\bf \theta}=\{{\bf w}, {\bf b}\}$ denotes the parameters to be optimized in the neural network. Building upon the two training methods mentioned in the previous subsection, we define the following two loss functions:
\begin{equation}\label{loss}
  {\rm MSE_1}({\bf \theta})=\frac{1}{N_T N}\sum_{i=0}^{N_T-1}\sum_{j=0}^{N-1}\left|\mathcal{R}(n^j_{f_1}, t^i_f; {\bf \theta})\right|^2, \ \ {\rm MSE_2}({\bf \theta})=\frac{1}{N_T N_n}\sum_{i=0}^{N_T-1}\sum_{j=0}^{N_n-1}\left|\mathcal{R}(n^j_{f_2}, t^i_f; {\bf \theta})\right|^2,
\end{equation}
where the independent variables $n^j_{f_1}\in\Omega$, $n^j_{f_2}\in\Omega'$, and $t^i_f\in[t_0, t_0+i\frac{t_T-t_0}{N_T-1}] (i=1,2,...,N_T-1)$. By minimizing the mean squared error criterion, we optimize the parameters of the neural network to satisfy the provided physical information. We choose to use the Adam optimization algorithm to optimize the loss function. Ultimately, the trained PG-PhyCRNet model can not only provide numerical solutions within the specified region but also forecast and provide numerical solutions outside the time domain except for the boundary values.

\begin{figure*}[!htb]
    \begin{center}
    {\scalebox{0.7}[0.7]{\includegraphics{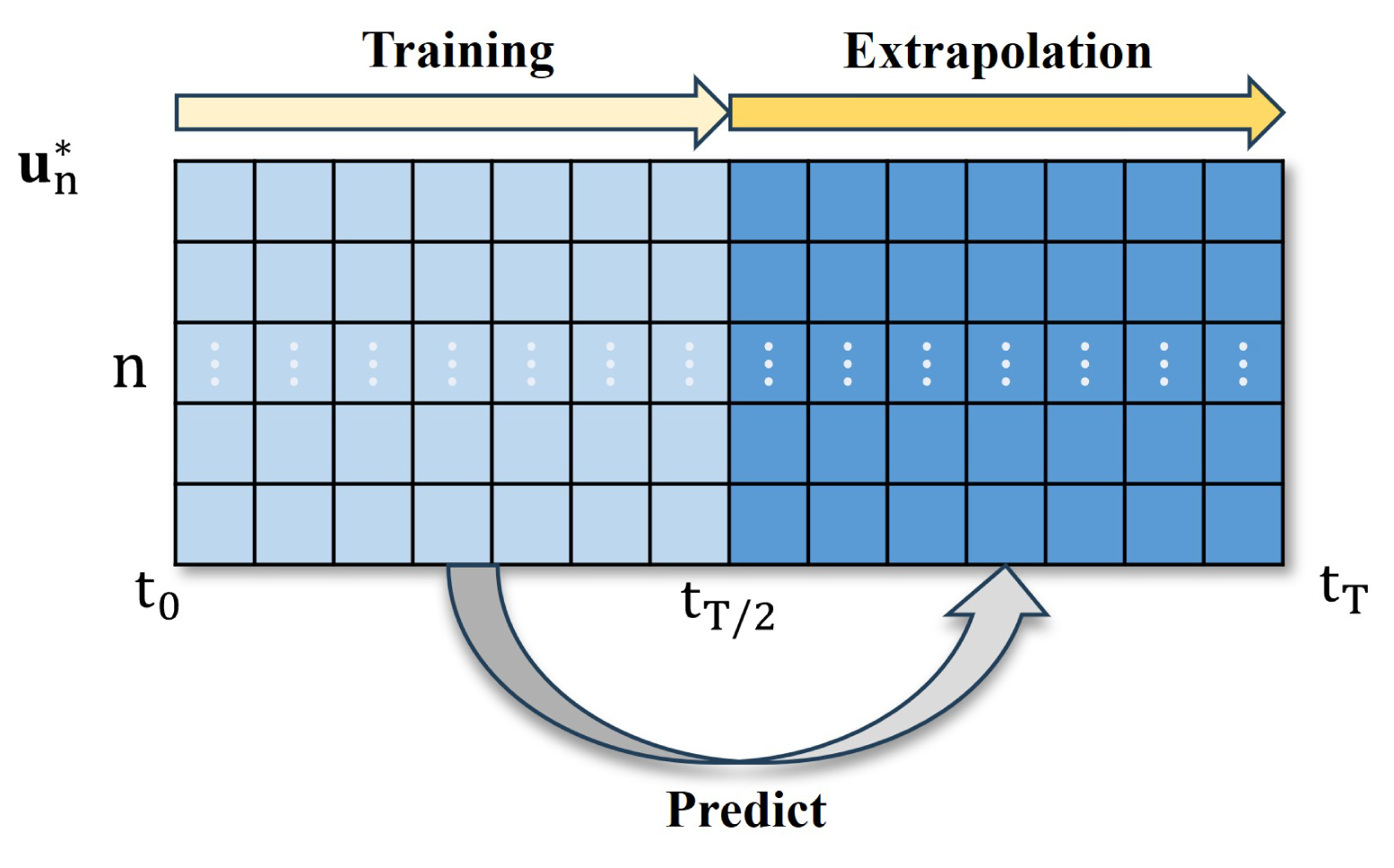}}}
    \end{center}
    \vspace{-0.15in} \caption{\small The schematic diagram of the internal region training and extrapolation prediction.} \label{extrapolation}
\end{figure*}

We need to use the exact solution of the physical equations to evaluate the application performance of PG-PhyCRNet in discrete integrable systems and assess the propagation error through two stages: training and extrapolation. Although the use of PhyCRNet involves the above two methods to meet different needs, the ultimate goal is to evaluate the predictive accuracy of the model on integer grid sites. The first stage involves evaluating the model's training generalization ability through the relative $\mathbb{L}_2$ error of $N\times N_t$ ($N_t=(N_T+1)/2$) data points on the given computational grid $\Omega\times[t_0, t_0+i\frac{t_{T/2}-t_0}{N_t-1}] (i=1,2,...,N_t-1)$:
\begin{equation}
    RE_t=\frac{\sqrt{\sum_{k=0}^{N-1}\sum_{l=0}^{N_t-1}\left|\widehat{u}^{k,l}-u\left(n_{k+1}, t_0+l\frac{t_{T/2}-t_0}{N_t-1}\right)\right|^2}}{\sqrt{\sum_{k=0}^{N-1}\sum_{l=0}^{N_t-1}\left|u\left(n_{k+1}, t_0+l\frac{t_{T/2}-t_0}{N_t-1}\right)\right|^2}},
\end{equation}
the second stage involves evaluating the model's extrapolation generalization ability through the relative $\mathbb{L}_2$ error of $N\times N_t'$ ($N_t'=N_t$) data points on the given computational grid $\Omega\times[t_{T/2}, t_{T/2}+i\frac{t_T-t_{T/2}}{N_t'-1}] (i=1,2,...,N_t'-1)$:
\begin{equation}
    RE_e=\frac{\sqrt{\sum_{k=0}^{N-1}\sum_{l=0}^{N_t'-1}\left|\widehat{u}^{k,N_t+l-1}-u\left(n_{k+1}, t_1+l\frac{t_T-t_{T/2}}{N_t'-1}\right)\right|^2}}{\sqrt{\sum_{k=0}^{N-1}\sum_{l=0}^{N_t'-1}\left|u\left(n_{k+1}, t_1+l\frac{t_T-t_{T/2}}{N_t'-1}\right)\right|^2}},
\end{equation}
where $\widehat{u}$ and $u\left(\cdot\right)$ stand for predictive value and true value respectively. Fig.~\ref{extrapolation} illustrates the schematic diagram of the internal region training and extrapolation prediction. Fig.~\ref{extrapolation} clearly shows that the model trained through iterations within the time domain $t\in[t_0,t_{T/2}]$ can be used for extrapolation predictions in the time domain $t\in[t_{T/2}, t_T]$.

\subsection{Network settings}
For the numerical results of any lattice equation presented below, we adopted the following network settings: in the encoder structure, the number of convolutional kernels in the three one-dimensional convolutional layers is ${16, 64, 256}$, the number of channels in the first convolution operation is 2 (all discussed are coupled lattice equations), and the kernel size of the three convolution operations is 4, with a stride of 2 and zero-padding applied. A ConvLSTM layer is applied on the latent space with 128 nodes/cell/hidden states, with kernel sizes of 3, a stride of 1, and zero-padding applied. The decoder consists of a pixel shuffle operation with an upsampling factor of 8 and a convolution operation with a stride of 1, including two convolution kernels of size 5, with the number of channels matching the corresponding number of channels after the pixel shuffle operation. We chose the time steps in the interval $[t_0, t_{T/2}]$ for training and internal prediction, and in the interval $[t_{T/2}, t_T]$ for extrapolation prediction. The learning rate was set to $1\times10^{-3}$. 

All code was executed in an environment with Python $3.9$ and PyTorch $1.13$. The numerical experiments were conducted on a professional computer equipped with GPU NVIDIA GeForce RTX $3090$. 

\section{Numerical results without pseudo grid training}\label{sec3}
In this section, our aim is to obtain high-precision numerical results for integer grid points within a given domain and reliable extrapolated results beyond their time domain. To achieve this, we consider training with initial condition data from $n\in\mathbb{Z}$, and demonstrate this result using the localized wave solutions of two classical lattice equations. 

\subsection{One-soliton solution of Toda lattice}
The Toda equation describes a one-dimensional chain of particles, where each particle with a mass of $1$ interacts with its nearest neighbors through an exponential potential. We consider the initial boundary value problem for one-dimensional Toda lattice with coupled form~\cite{NLE1}:
\begin{equation}\label{toda}
  \begin{cases}
    \dot{u}_n = u_n(v_n-v_{n-1}),\\
    \dot{v}_n = u_{n+1}-u_n,\ \ \ n\in\Omega, \ t\in[t_0, t_T], \\
    u_{n}(t_0)=u(n,t_0), \ v_{n}(t_0)=v(n,t_0), \\
    u_{n_1}= u_{b_1}, \ u_{n_N}= u_{b_2},\\ 
    v_{n_1}= v_{b_1}, \ v_{n_N}= v_{b_2}.
  \end{cases}
\end{equation}
This model finds applications in describing the dynamical properties of various systems such as LC transmission lines~\cite{NLE1}, biopolymers (e.g., DNA chains~\cite{DNA}), excitations in anharmonic lattices, and lattices of optical solitons in fibers~\cite{optik}. Besides, in the continuous limit (\ref{toda_cl}), Eq.~(\ref{toda}) approximates the KdV equation. Therefore, the discrete solutions of Eq.~(\ref{toda}) can be used to indirectly explore the dynamical behavior of the solutions to the KdV equation. One soliton solution of Eq.~(\ref{toda}) is~\cite{todasoliton}:
\begin{equation}\label{1soliton-toda}
\begin{aligned}
  {u}_n &= \frac{\cosh\left[q_0\left(n-n_0(t)\right)\right]\left\{\sinh(q_0)e^{q_0\left(n-n_0(t)-2\right)}+\cosh\left[q_0\left(n-n_0(t)-1\right)\right]\right\}e^{-q_0}}{\cosh\left[q_0\left(n-n_0(t)-1\right)\right]^2},\\
  {v}_n &= \frac{\sinh\left(q_0\right)^2}{\cosh\left[q_0\left(n-n_0(t)-1\right)\right]\cosh\left[q_0\left(n-n_0(t)\right)\right]},
\end{aligned}
\end{equation}
where $n_0(t)=n_0-\sinh(q_0)/q_0t$. In this case, let $q_0=1/2$, $n_0=0$, $\Omega=[-12, 13]$ is a
finite one-dimension lattice and $[t_0, t_1]=[-0.5, 0.5]$. Substituting these conditions into Eq.~(\ref{1soliton-toda}), we can obtain the initial value for Eq.~(\ref{toda}). In addition, we default the boundary values of the potential functions $u_n$ and $v_n$ (in the constant background) to $u_{b_1}=u_{b_2}=1$ and $v_{b_1}=v_{b_2}=0$. A time step size of $\delta_t=0.005$ was chosen in the time domain to account for the approximation of derivatives in the lattice equation and the implicit time progression in ConvLSTM. Therefore, the region we use as training is $R_{in} (R_{in}=[n,t], n\in[-11,12], t\in[-0.5, 0])$ and the training was conducted with $100$ time steps per epoch in the time interval $t\in[-0.5, 0]$. Additionally, predictions are made in region $R_{out} (R_{out}=[n,t], n\in[-11,12], t\in[0, 0.5])$ to assess the extrapolation capability of the trained model.

After training for $2000$ epochs, PhyCRNet achieved a significant reduction in the loss function, reaching $2.320E-08$ after $1003$ seconds. Subsequently, we conducted predictions in the time domain $t\in[-0.5, 0]$ and obtained excellent numerical results, in which the $\mathbb{L}_2$ error of component $u_n$ reaches $2.0567724E-05$, and that of component $v_n$ reaches $2.2417906E-04$. Moreover, extrapolation predictions in the time domain $t\in[0, 0.5]$ also yielded good numerical results, with the $\mathbb{L}_2$ error for component $u_n$ at $1.0692835E-03$, and for component $v_n$ at $1.3986321E-02$. 

\begin{figure*}[!htb]
    \begin{center}
    {\scalebox{0.32}[0.32]{\includegraphics{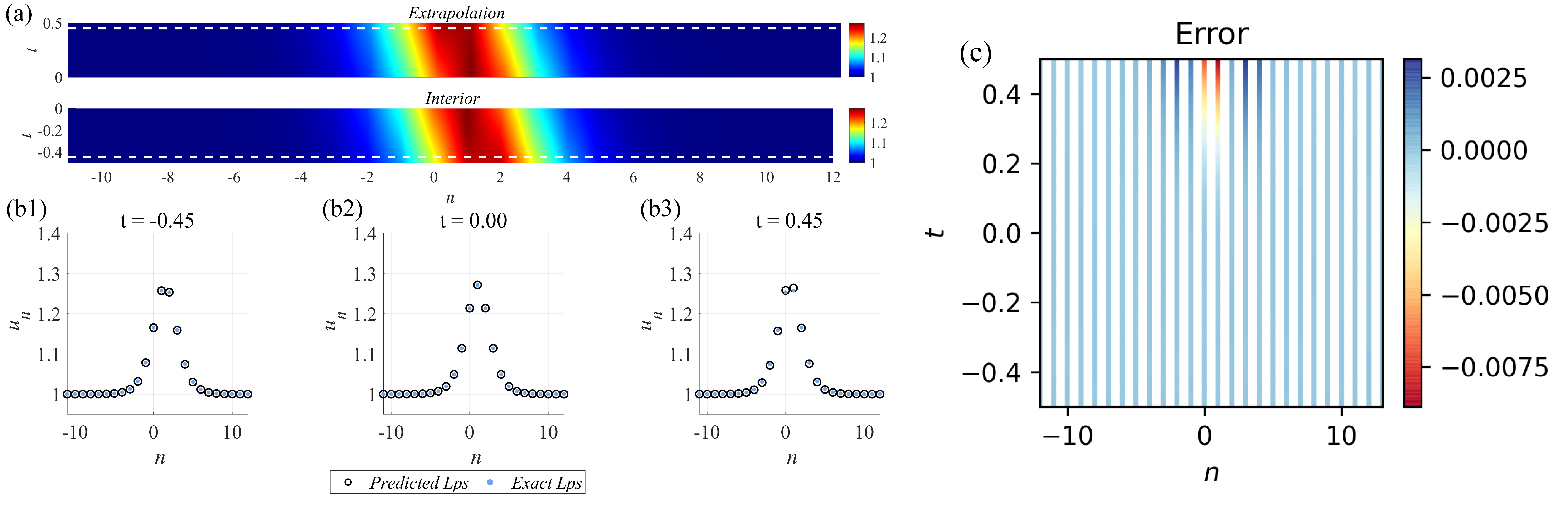}}}
    \end{center}
    \vspace{-0.15in} \caption{\small The numerical one-soliton solution of the Toda lattice for the component $u_n$ is presented. (a) illustrates the density plot of component $u_n$, with the lower sub-figure depicting the prediction within the time domain $t\in[-0.5, 0]$, and the upper sub-figure representing the extrapolated prediction within the time domain $t\in[0, 0.5]$. The white dashed lines indicate the selected moments at $t = -0.45$ and $t = 0.45$; (b1)-(b3) show the temporal evolution of one-soliton at three different moments, where "Lps" stands for lattice points. (c) presents the density plot of the error (predicted solution minus exact solution).} \label{toda1_u}
\end{figure*}

\begin{figure*}[!htb]
    \begin{center}
    {\scalebox{0.32}[0.32]{\includegraphics{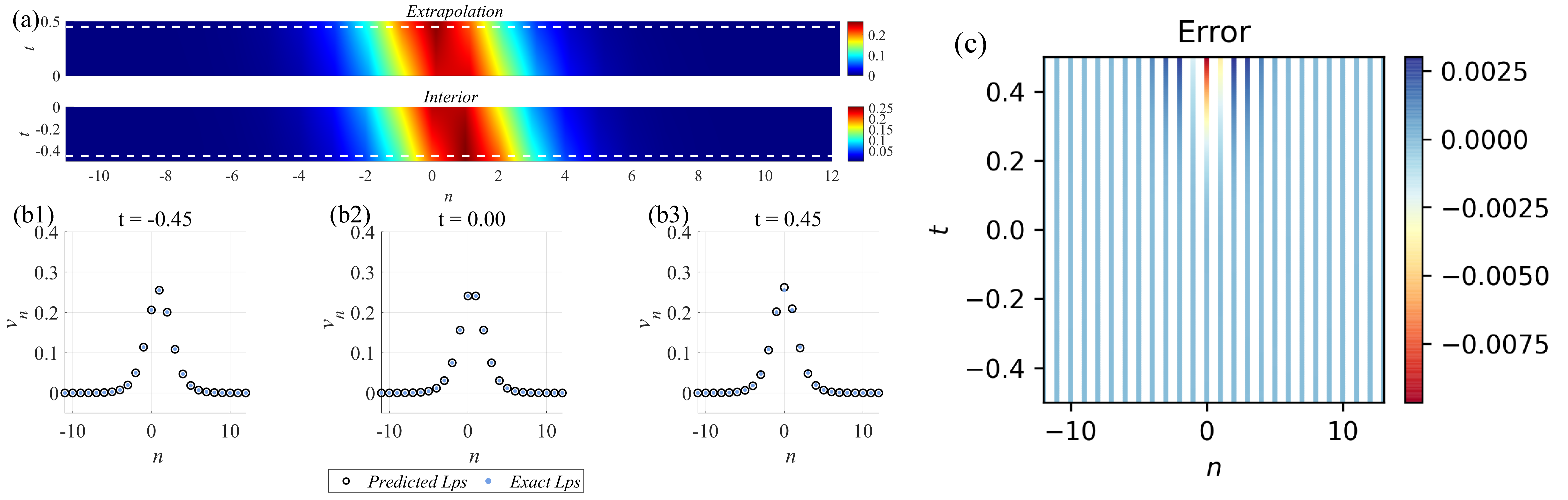}}}
    \end{center}
    \vspace{-0.15in} \caption{\small The numerical one-soliton solution of the Toda lattice for the component $v_n$ is presented. (a) illustrates the density plot of component $v_n$, with the lower sub-figure depicting the prediction within the time domain $t\in[-0.5, 0]$, and the upper sub-figure representing the extrapolated prediction within the time domain $t\in[0, 0.5]$. The white dashed lines indicate the selected moments at $t = -0.45$ and $t = 0.45$; (b1)-(b3) show the temporal evolution of one-soliton at three different moments. (c) presents the density plot of the error (predicted solution minus exact solution).} \label{toda1_v}
\end{figure*}

\begin{figure*}[!htb]
    \begin{center}
    {\scalebox{0.32}[0.32]{\includegraphics{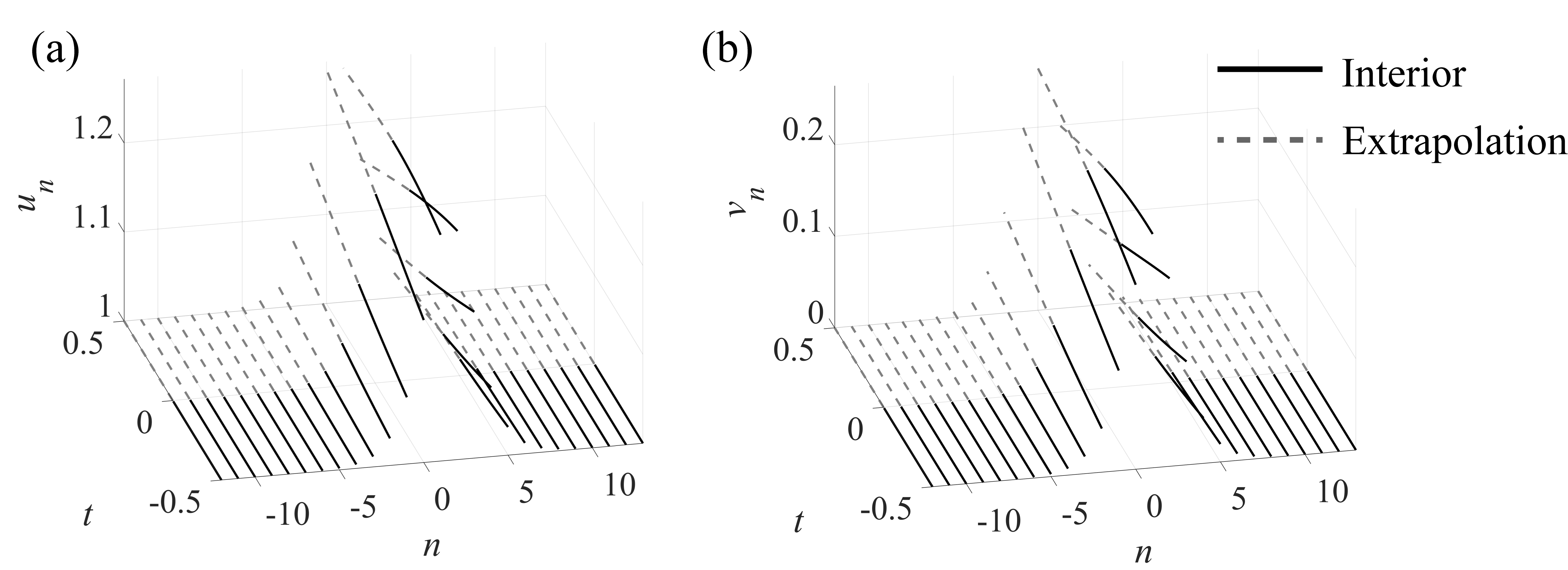}}}
    \end{center}
    \vspace{-0.15in} \caption{\small Three-dimensional plots of the numerical solution of one-soliton for Toda lattice, where the solid black line indicates the prediction within the time domain $t\in[-0.5, 0]$, and the grey dashed line indicates the extrapolated prediction within the time domain $t\in[0, 0.5]$.} \label{toda1_3d}
\end{figure*}

Figs.~\ref{toda1_u} and \ref{toda1_v} respectively present the density plots, cross-sectional plots (temporal evolution plots), and error dynamics plots of one-soliton solutions for Toda lattice based on the PhyCRNet method. The density plot in (a) illustrates the evolution of the line soliton along the negative direction of the $n$-axis, which is clearly evident in the figure. Additionally, the consistency between the predicted and exact solutions in the cross-sectional plots (b1)-(b2) indicates that the model has achieved good performance in learning the one-soliton solution within the time domain $t\in[-0.5, 0]$, while (b3) demonstrates successful extrapolation of the one-soliton solution within the time domain $t\in[0, 0.5]$. From (c), it can be observed that the model's generalization error mainly arises from the long-term evolution of the soliton, suggesting that the neural network's ability to transmit information at different time steps needs improvement, and this error partly originates from the differential results in the temporal direction. Furthermore, Fig. \ref{toda1_3d} shows the three-dimensional plots of the numerical solutions $u_n$ and $v_n$, providing a clearer depiction of the line soliton's dynamic behavior. It can be found in Table.~\ref{tab01} the neural network has successfully learned the dynamic behavior of the single soliton solution on the Toda lattice, as supported by the evidence presented above.
\begin{table}
\centering
\caption{Errors of learning the one-soliton solution for Toda lattice }
\label{tab01}
\begin{tblr}{
  hline{1,4} = {-}{0.1em},
  hline{2} = {-}{},
}
              & $u_n$            & $v_n$     & Total      \\
Interior      & 2.0567724E-05        & 2.2417906E-04 & 2.7090538E-05  \\
Extrapolation & 1.0692835E-03        & 1.3986321E-02 & 1.5350957E-03 
\end{tblr}
\end{table}

\subsection{Dark-bright rational soliton solution of self-dual network equation}
We investigate the self-dual network equation with the initial boundary value problem:
\begin{equation}\label{sd}
  \begin{cases}
    \dot{I}_n = (1+I_n^2)(V_{n-1}-V_n),\\
    \dot{V}_n = (1+V_n^2)(I_{n}-I_{n+1}),\ \ \ n\in\Omega, \ t\in[t_0, t_1], \\
    I_{n}(t_0)=I(n,t_0), \ V_{n}(t_0)=V(n,t_0), \\
    I_{n_1}= I_{b_1}, \ I_{n_N}= I_{b_2},\\ 
    V_{n_1}= V_{b_1}), \ V_{n_N}= V_{b_2}.
  \end{cases}
\end{equation}
The nonlinear self-dual network equation describes the propagation of electrical signals in a ladder-type electric circuit, where $V_n$ and $I_n$ represent the voltage and current~\cite{hirota_sd}. Besides, in the continuous limit (\ref{sd_cl}), Eq.~(\ref{sd}) approximates the mKdV equation. Therefore, the discrete solutions of Eq.~(\ref{sd}) can be used to indirectly explore the dynamical behavior of the solutions to the mKdV equation. Based on the Lax pair of Eq.~(\ref{sd})~\cite{NLE2}
\begin{equation}\label{sd_lax}
  E\psi_{n}=U_n\psi_n=\left(\begin{array}{cc}
                      \lambda+I_nV_n & -I_n+V_n/\lambda \\
                      I_n-\lambda V_n & 1/\lambda+I_nV_n 
                    \end{array}\right)\psi_n, \ \ 
                    \dot{\psi}_n=W_n\psi_n=\left(\begin{array}{cc}
                      \lambda-I_nV_{n-1} & -I_n+V_{n-1}/\lambda \\
                      I_n-\lambda V_{n-1} & 1/\lambda-I_nV_{n-1} 
                    \end{array}\right)\psi_n,
\end{equation}
where $E$ stands for shift operator, $\psi_n(t)=(\psi_{1,n}, \psi_{2,n})$ is the vector eigenfunction, $\lambda$ is the eigenvalue parameter, the Darboux matrix $T_n(t; \lambda)$ can be defined as\cite{DT3}:
\begin{equation}\label{dt_m}
  T_n(t; \lambda)=\left(\begin{array}{cc}
                    \lambda^N+\sum_{j=0}^{N-1}A_n^{(j)}\lambda^j & \sum_{j=0}^{N-1}B_n^{(j)}\lambda^j \\
                    -\lambda^N\sum_{j=0}^{N-1}B_n^{(j)}\lambda^{-j} & 1+\lambda^N \sum_{j=0}^{N-1}A_n^{(j)}\lambda^{-j}
                  \end{array}\right),
\end{equation}
where $A_n^{(j)}$ and $B_n^{(j)}$ can be determined by $2N$ linear system of equations obtained by Taylor expanding $\lim_{\epsilon\rightarrow0}T(t;\lambda_s+\epsilon)\psi_n(t;\lambda_s+\epsilon)/\epsilon^k_s=0 (s=1,2,...M; k_s=0,1,...,v_s, e.g. N=M+\sum_{s=1}^{M}v_s)$ at $\epsilon=0$. Hence, the $N$-order rational soliton solution can be obtained by Darboux transformation of Eq.~(\ref{sd}):
\begin{equation}\label{dt_sd}
  \widehat{I}^{(N)}_n = \frac{I_n+B_n^{(N-1)}}{A_n^{(0)}}, \ \ \widehat{V}^{(N)}_n=B_{n+1}^{(0)}+V_nA_{n+1}^{(0)}.
\end{equation}
Let $N=1$, the unknown functions $A_n^{(0)}=\delta A_n^{(0)}/\delta_n$ and $B_{n}^{(0)}=\delta B_{n}^{(0)}/\delta_n$ are
\begin{equation}\label{ab0}
  \delta A_n^{(0)}=\left|\begin{array}{cc}
  -\lambda_1\psi^{(0)}_{1,n} & \psi^{(0)}_{2,n} \\
  -\psi^{(0)}_{2,n} & -\lambda_1\psi^{(0)}_{1,n}
  \end{array}\right|, \ \ \delta B_{n}^{(0)}=\left|\begin{array}{cc}
                            \psi^{(0)}_{1,n} & -\lambda_1\psi^{(0)}_{1,n} \\
                            \lambda_1\psi^{(0)}_{2,n} & -\psi^{(0)}_{2,n}
                             \end{array}\right|, \ \ \delta_{n}=\left|\begin{array}{cc}
                            \psi^{(0)}_{1,n} & \psi^{(0)}_{2,n} \\
                            \lambda_1\psi^{(0)}_{2,n} & -\lambda_1\psi^{(0)}_{1,n}
                             \end{array}\right|,
\end{equation}
where the non-zero seed solutions $I_n=\alpha=\frac{11}{60}$, $V_n=0$. The vector eigenfunction of Eq.~(\ref{sd_lax}) $\psi_n^{(0)}=(\psi_{1,n}^{(0)}, \psi_{2,n}^{(0)})$ can be determined by
\begin{equation}\label{taylor}
  \psi_n(t;\lambda,\epsilon)|_{\lambda=\lambda_1+\epsilon} = \left(\begin{array}{c}
                                                               \psi_{1,n}^{(0)} \\
                                                               \psi_{2,n}^{(0)} 
                                                             \end{array} \right) +\left(\begin{array}{c}
                                                               \psi_{1,n}^{(1)} \\
                                                               \psi_{2,n}^{(1)} 
                                                             \end{array} \right)\epsilon+\left(\begin{array}{c}
                                                               \psi_{1,n}^{(2)} \\
                                                               \psi_{2,n}^{(2)} 
                                                             \end{array} \right)\epsilon^2+..., \ \ \lambda_1=\alpha+\sqrt{1+\alpha^2},
\end{equation}
with
\begin{equation}\label{ra}
  \psi_{1,n}^{(0)}=\beta_n(t)\frac{671t + 660n + 3660}{60390}, \ \ \psi_{2,n}^{(0)}=\beta_n(t)\frac{61t + 60n}{5490}, \ \ \beta(n,t)=\frac{61^{n+\frac{1}{2}}\sqrt{11}}{15^{n-\frac{1}{2}}4^n}e^{\frac{61}{60}t}.
\end{equation}
Substituting Eq.~(\ref{ra}) into (\ref{ab0}) (\ref{dt_sd}) and eliminating $\beta(n,t)$, we find that the mathematical expression for a rational soliton is entirely composed of rational polynomials. This is in contrast to traditional soliton expressions, which involve hyperbolic functions (in Eq.~(\ref{1soliton-toda})). Of particular interest is whether the PhyCRNet method can effectively approximate this function and explore its dynamic behavior. 

In this case, let $\Omega=[-31, 26]$ is a finite one-dimension lattice and $[t_0, t_1]=[-1, 1]$. The initial value for Eq.~(\ref{toda}) can be obtained by Eq.~(\ref{dt_sd}). In addition, we default the boundary values of the potential functions $I_n$ and $V_n$ (in the constant background) to $I_{b_1}=I_{b_2}=0$ and $V_{b_1}=V_{b_2}=-\alpha$. A time step size of $\delta_t=0.01$ was chosen in the time domain to account for the approximation of derivatives in the lattice equation and the implicit time progression in ConvLSTM. Therefore, the region we use as training is $R_{in} (R_{in}=[n,t], n\in[-30,25], t\in[-1, 0])$ and the training was conducted with $100$ time steps per epoch in the time interval $t\in[-1, 0]$. Additionally, predictions are made in region $R_{out} (R_{out}=[n,t], n\in[-30,25], t\in[0, 1])$ to assess the extrapolation capability of the trained model. After training for $2000$ epochs, PhyCRNet achieved a significant reduction in the loss function, reaching the order of $5.951E-07$ after $1138$ seconds. Subsequently, we conducted predictions in the time domain $t\in[-1, 0]$ and obtained excellent numerical results, in which the $\mathbb{L}_2$ error of component $I_n$ reaches $8.877356E-04$, and that of component $V_n$ reaches $9.527011E-04$. Moreover, extrapolation predictions in the time domain $t\in[0, 1]$ also yielded good numerical results, with the $\mathbb{L}_2$ error for component $I_n$ at $3.014642E-02$, and for component $V_n$ at $2.244632E-02$.

\begin{figure*}[!htb]
    \begin{center}
    {\scalebox{0.32}[0.32]{\includegraphics{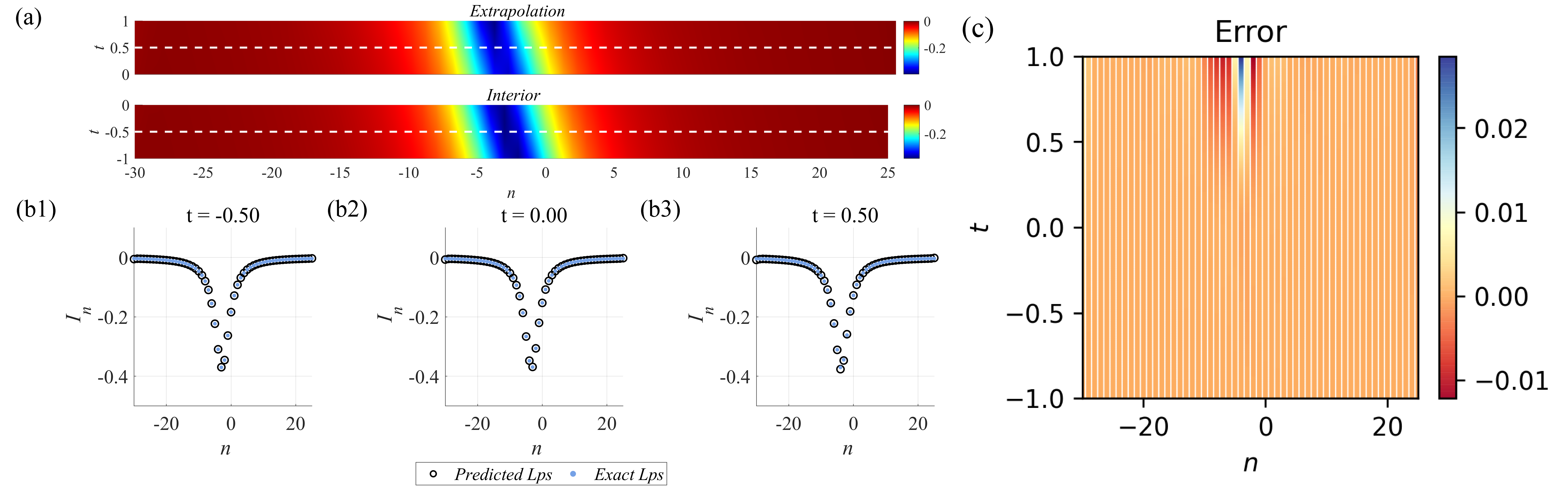}}}
    \end{center}
    \vspace{-0.15in} \caption{\small The numerical dark rational soliton solution of the self-dual network equation for the component $I_n$ is presented. (a) illustrates the density plot of component $I_n$, with the lower sub-figure depicting the prediction within the time domain $t\in[-1, 0]$, and the upper sub-figure representing the extrapolated prediction within the time domain $t\in[0, 1]$. The white dashed lines indicate the selected moments at $t = -0.5$ and $t = 0.5$; (b1)-(b3) show the temporal evolution of rational soliton solution at three different moments. (c) presents the density plot of the error (predicted solution minus exact solution).} \label{sd_r1_u}
\end{figure*}

\begin{figure*}[!htb]
    \begin{center}
    {\scalebox{0.32}[0.32]{\includegraphics{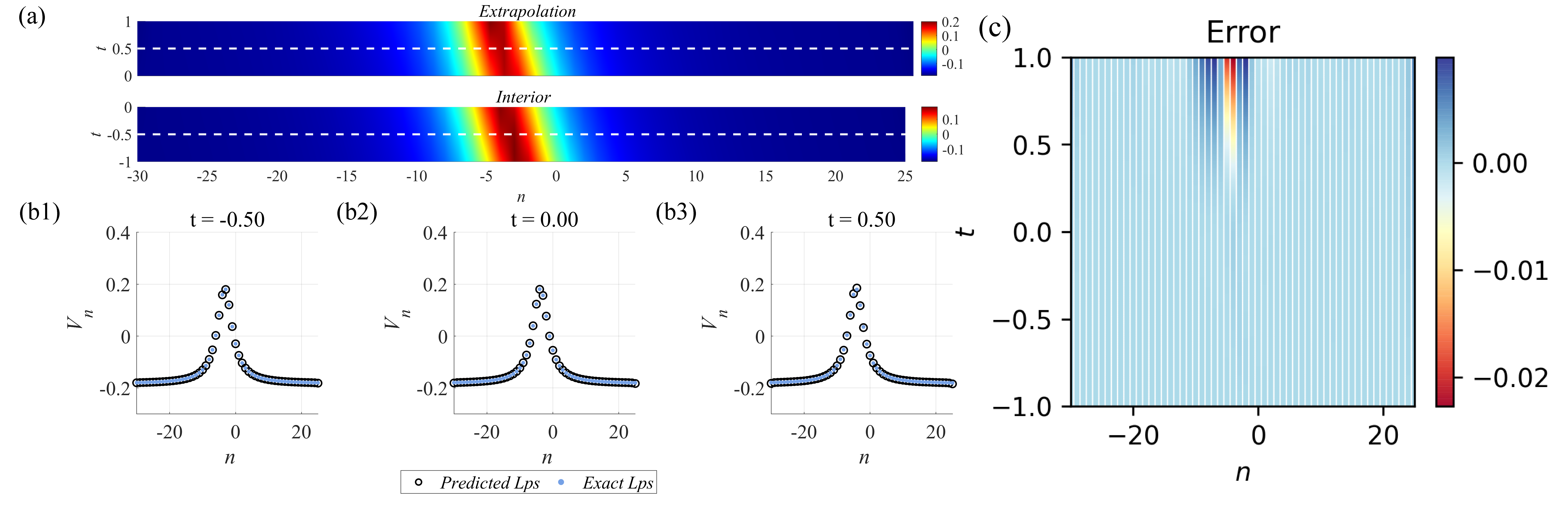}}}
    \end{center}
    \vspace{-0.15in} \caption{\small The numerical bright rational soliton solution of the self-dual network equation for the component $V_n$ is presented. (a) illustrates the density plot of component $V_n$, with the lower sub-figure depicting the prediction within the time domain $t\in[-1, 0]$, and the upper sub-figure representing the extrapolated prediction within the time domain $t\in[0, 1]$. The white dashed lines indicate the selected moments at $t = -0.5$ and $t = 0.5$; (b1)-(b3) show the temporal evolution of rational soliton solution at three different moments. (c) presents the density plot of the error (predicted solution minus exact solution).} \label{sd_r1_v}
\end{figure*}

Figs.~\ref{sd_r1_u} and \ref{sd_r1_v} respectively present the density plots, cross-sectional plots (temporal evolution plots), and error dynamics plots of dark-bright rational soliton solutions for self-dual network equation based on the PhyCRNet method. The density plot in (a) illustrates the evolution of the rational soliton along the negative direction of the $n$-axis, which is clearly evident in the figure. Additionally, the consistency between the predicted and exact solutions in the cross-sectional plots (b1)-(b2) indicates that the model has achieved good performance in learning the rational soliton solution within the time domain $t\in[-1, 0]$, while (b3) demonstrates successful extrapolation of the rational soliton solution within the time domain $t\in[0, 1]$. In panel (c), errors arising from the propagation of rational soliton solutions over time can also be observed. Furthermore, Fig. \ref{toda1_3d} shows the three-dimensional plots of the numerical solutions $I_n$ and $V_n$, providing a clearer depiction of the dark-bright rational soliton's dynamic behavior. It can be found in Table.~\ref{tab02} the neural network has successfully learned the dynamic behavior of the rational soliton solution on the self-dual network equation, as supported by the evidence presented above. 
\begin{figure*}[!htb]
    \begin{center}
    {\scalebox{0.32}[0.32]{\includegraphics{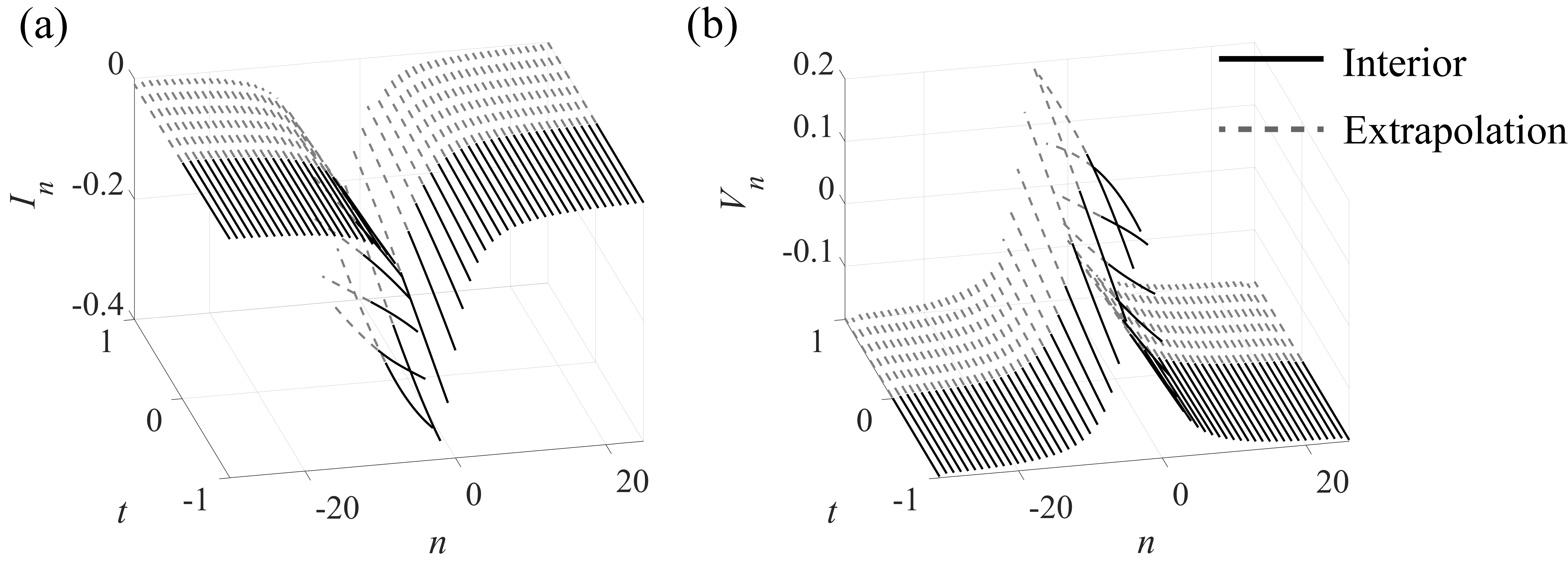}}}
    \end{center}
    \vspace{-0.15in} \caption{\small Three-dimensional plots of the numerical solution of dark-bright rational soliton for self-dual network equation, where the solid black line indicates the prediction within the time domain $t\in[-1, 0]$, and the grey dashed line indicates the extrapolated prediction within the time domain $t\in[0, 1]$.} \label{sd_r1_3d}
\end{figure*}
\begin{table}
\centering
\caption{Errors of learning the first-order rational solution for self-dual network equation}
\label{tab02}
\begin{tblr}{
  hline{1,4} = {-}{0.1em},
  hline{2} = {-}{},
}
              & $u_n$            & $v_n$     & Total      \\
Interior      & 8.877356E-04        & 9.527011E-04 & 9.160831E-04  \\
Extrapolation & 3.014642E-02        & 2.244632E-02 & 2.1315258E-02 
\end{tblr}
\end{table}

\section{Numerical results with pseudo grid training}\label{sec4}
After conducting several repeated experiments, it was observed that the model exhibited slight deficiencies in its extrapolation capabilities when dealing with soliton solutions characterized by higher wave speeds, narrower widths, or other complex features. Motivated by this observation, we aim for PG-PhyCRNet to effectively capture the evolution of solitons over time (i.e., its extrapolation capability in the temporal direction) in the training of soliton solutions with higher wave speeds or complexity. This may involve sacrificing a small portion of the training error within the given spatiotemporal region. In this section, we will continue to discuss different complex solitary waves using these two lattice models. The introduction of pseudo grid necessitates partitioning the spatial domain grid in Eq.~(\ref{toda}) and~(\ref{sd}) as $\Omega = [n_1, n_1+i\frac{n_N-n_1}{N_n-1}],\ i=1,...,N_n-1$. To investigate the effects of pseudo grid on the PG-PhyCRNet framework, we evaluated the performance differences between PhyCRNet with and without pseudo grid for the cases of two solitons in the Toda lattice and one- or two- solitons in the self-dual network equation.

\subsection{Two-soliton solution of Toda lattice under the non-zero background}
In this section's numerical experiments, we will continue to discuss the initial-boundary value problem~(\ref{toda}) for the one-dimensional Toda lattice. Based on the Darboux transformation in Ref.~\cite{wen_toda}, we give the two-soliton solution of Toda lattice under the non-zero background:
\begin{equation}\label{toda_dt}
  \widehat{u}_n = u_n-a_n+a_{n+1}, \ \ \widehat{v}_n=\frac{v_n+b_n}{1+c_n}.
\end{equation}
where
\begin{equation*}
\begin{aligned}
  &a_n=\frac{\lambda_2\delta_{1,n}-\lambda_1\delta_{2,n}}{\delta_{2,n}-\delta_{1,n}}, \ b_n = \frac{\lambda_1-\lambda_2}{\delta_{2,n}-\delta_{1,n}}, \ c_n = \frac{(\lambda_2-\lambda_1)\delta_{1,n}\delta_{2,n}}{(\lambda_1-\lambda_2)\delta_{1,n}\delta_{2,n}+\delta_{2,n}-\delta_{1,n}}, \\ &\delta_{i,n}=\frac{r_i\tau_-(\lambda_i)^{n}e^{\rho_-(\lambda_i)t}-\tau_+(\lambda_i)^{n}e^{\rho_+(\lambda_i)t}}{\tau_+(\lambda_i)^{n}e^{\rho_+(\lambda_i)t}-r_i\tau_-(\lambda_i)^{n+1}e^{\rho_-(\lambda_i)t}}, \ \tau_{\pm}(\lambda_i)=1+\lambda_i\pm\frac{1}{2}\sqrt{\lambda_i^2+2\lambda_i-3}, \ \rho(\lambda_i)_{\pm}=\frac{\lambda_i\tau{\pm}-2}{2\tau_{\pm}} (i=1,2).
\end{aligned}
\end{equation*}
In this case, let $r_1=-1$, $r_2=1$, $\lambda_{1}=2$, $\lambda_2=2.1$, the spatial domain $\Omega=[-12, 13]$ is divided into $225$ segments $(n_1=-12, n_N=13, \delta_n=\frac{1}{9})$, while the time domain $[t_0, t_1]=[-1, 1]$ is divided into $200$ segments (with a training time step of $100$ and the remaining $100$ segments used for extrapolation predictions). Substituting these conditions into Eq.~(\ref{toda_dt}), we can obtain the initial boundary value problem for Eq.~(\ref{toda}). In addition, we default the boundary values of the potential functions $u_n$ and $v_n$ (in the constant background) to $u_{b_1}=u_{b_2}=1$ and $v_{b_1}=v_{b_2}=-1$. A time step size of $\delta_t=0.005$ was chosen in the time domain to account for the approximation of derivatives in the lattice and the implicit time progression in ConvLSTM. Therefore, the region we use as training is $R_{in} (R_{in}=[n,t], n\in[-12+\delta_n,13-\delta_n], t\in[-1, 0])$ and the training was conducted with $100$ time steps per epoch in the time interval $t\in[-1, 0]$. Additionally, the predictions are made in region $R_{out} (R_{out}=[n,t], n\in[-12+\delta_n,13-\delta_n], t\in[0, 1])$ to assess the extrapolation capability of the trained model.
After training for $1000$ epochs, the loss reaches $2.446500E-06$ after $688$ seconds. Similarly, training and predictions were conducted in the time domain $t\in[-1, 0]$. The $\mathbb{L}_2$ error for 
component $u_n$ reaches $1.462330E-03$, while that for component $v_n$ reaches $1.363670E-03$. We further conducted extrapolation predictions for the next $100$ time steps. The $\mathbb{L}_2$ error of component $u_n$ in the time domain $t\in[0, 1]$ reaches $2.730460E-03$, and for component $u_n$ reaches $2.687530E-03$.

\begin{figure*}[!htb]
    \begin{center}
    {\scalebox{0.32}[0.32]{\includegraphics{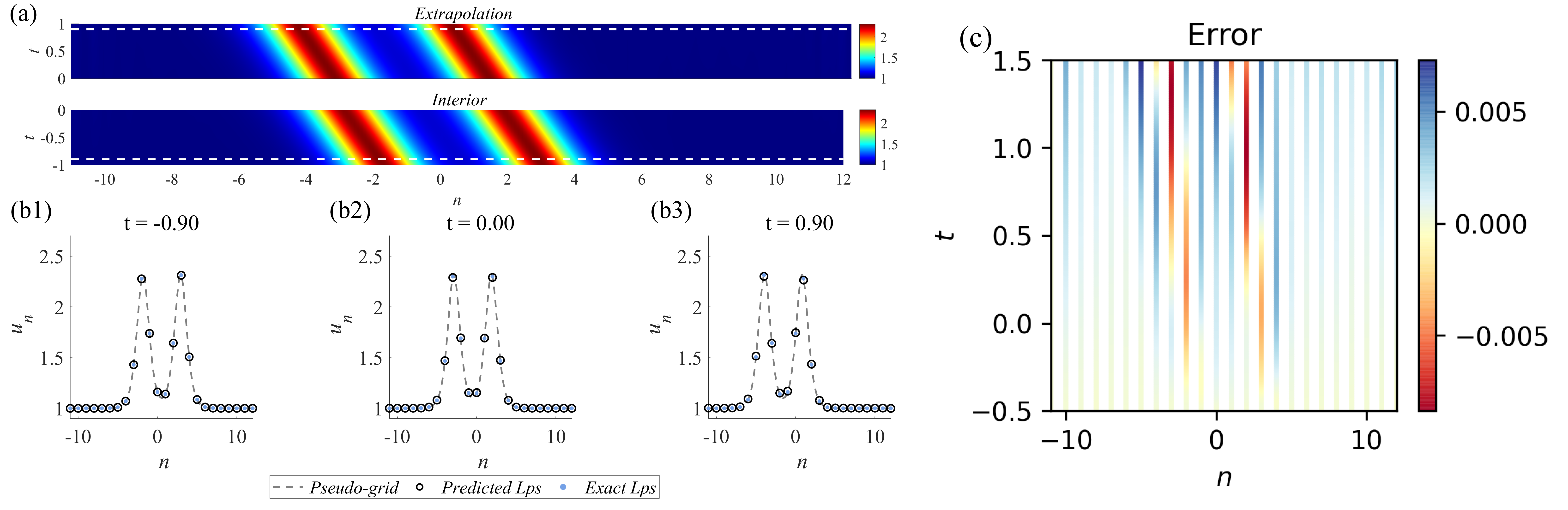}}}
    \end{center}
    \vspace{-0.15in} \caption{\small The numerical two-soliton solution of Toda lattice for the component $u_n$ is presented. (a) illustrates the density plot of component $u_n$, with the lower sub-figure depicting the prediction within the time domain $t\in[-1, 0]$, and the upper sub-figure representing the extrapolated prediction within the time domain $t\in[0, 1]$. The white dashed lines indicate the selected moments at $t = -0.9$ and $t = 0.9$; (b1)-(b3) show the temporal evolution of two-soliton solution at three different moments. (c) presents the density plot of the lattice error (predicted solution minus exact solution).} \label{toda2_u}
\end{figure*}

\begin{figure*}[!htb]
    \begin{center}
    {\scalebox{0.32}[0.32]{\includegraphics{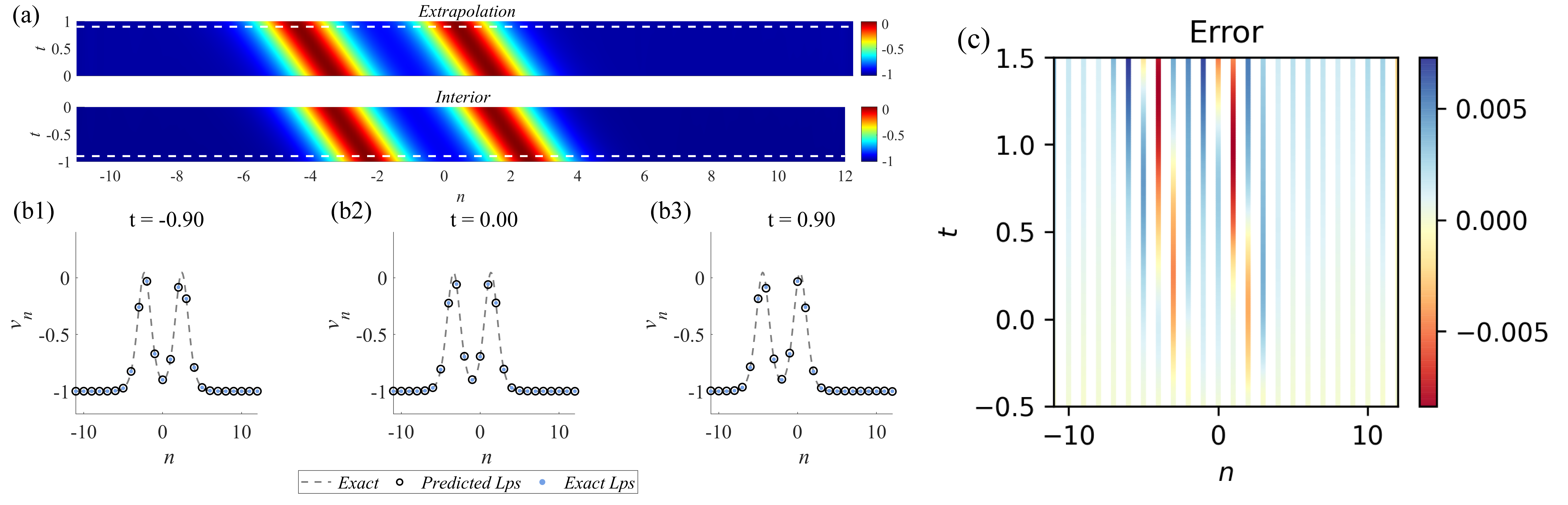}}}
    \end{center}
    \vspace{-0.15in} \caption{\small The numerical two-soliton solution of Toda lattice for the component $v_n$ is presented. (a) illustrates the density plot of component $v_n$, with the lower sub-figure depicting the prediction within the time domain $t\in[-1, 0]$, and the upper sub-figure representing the extrapolated prediction within the time domain $t\in[0, 1]$. The white dashed lines indicate the selected moments at $t = -0.9$ and $t = 0.9$; (b1)-(b3) show the temporal evolution of two-soliton solution at three different moments. (c) presents the density plot of the lattice error (predicted solution minus exact solution).} \label{toda2_v}
\end{figure*}

\begin{figure*}[!htb]
    \begin{center}
    {\scalebox{0.32}[0.32]{\includegraphics{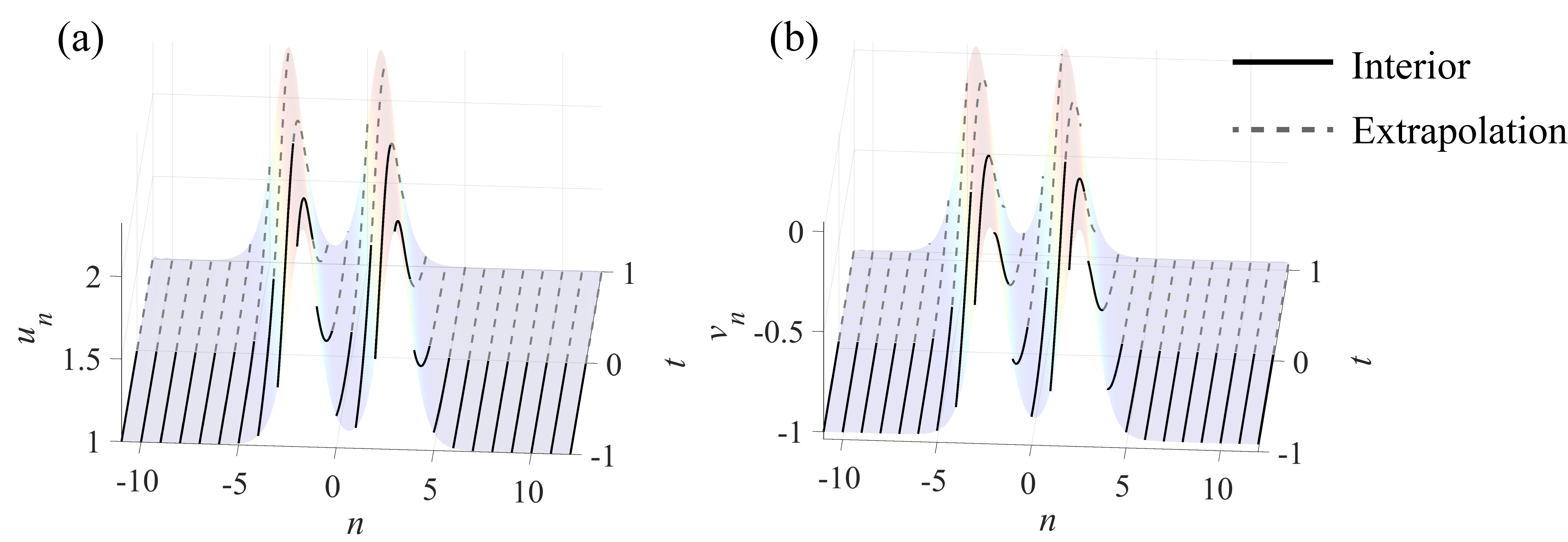}}}
    \end{center}
    \vspace{-0.15in} \caption{\small Three-dimensional plots of the numerical solution of two-soliton for Toda lattice, where the solid black line indicates the prediction within the time domain $t\in[-1, 0]$, and the grey dashed line indicates the extrapolated prediction within the time domain $t\in[0, 1]$. Colored surfaces represent predicted solutions with pseudo grid.} \label{toda2_3d}
\end{figure*}

Figs. \ref{toda2_u} and \ref{toda2_v} respectively present the density plots, temporal evolution plots, and error dynamics plots of two-soliton solutions for Toda lattice based on the PG-PhyCRNet method. From (a), it can be observed that two nearly parallel solitons move towards the negative direction of space as time evolves. Figs. (b1)-(b3) respectively show the time evolution of lattice points with the assistance of pseudo grid (grey dashed lines), where (b1)-(b2) represent predictions within the training region, and (b3) represents extrapolation results outside the time domain. The consistency between the predicted and exact solutions indicates that the model has achieved good performance in learning the two-soliton solution. Furthermore, Fig. \ref{toda2_3d} shows the three-dimensional plots of the numerical solutions $u_n$ and $v_n$. With an overall average generalization error of $1.431510E-03$ for predictions within the time domain $t\in[-1, 0]$ and $2.716800E-03$ for extrapolation outside the training range $t\in[0, 1]$, the neural network has successfully learned the dynamic behavior of two-soliton solution on the Toda lattice, as supported by the evidence presented above. The prediction accuracy for extrapolation is comparable to that within the region and does not exhibit significant differences.

Additionally, Table.~\ref{tab1} compares the PG-PhyCRNet model's performance in learning the two-soliton solution of the Toda lattice with and without the assistance of pseudo grid. It is evident that while the PG-PhyCRNet model without grid assistance achieves higher precision in predicting the two-soliton solution within the training region, its accuracy in extrapolation beyond this region is unsatisfactory. Conversely, the PG-PhyCRNet model with grid assistance, although sacrificing some precision in learning the two-soliton solution within the specified region, maintains a highly reliable extrapolation capability outside this region.
\begin{table}
\centering
\caption{Errors of two-soliton solution for Toda lattice with or without the assistance of pseudo grid training (1000 epoch)}
\label{tab1}
\begin{tblr}{
  cell{1}{2} = {c=3}{c},
  cell{1}{5} = {c=3}{c},
  hline{1,5} = {-}{0.1em},
  hline{2} = {2-7}{},
  hline{3} = {-}{},
}
              & Without~pseudo grid (670s for training time) &              &              & With~pseudo grid (688s for training time)&              &              \\
              & $u_n$            & $v_n$     & Total        & $u_n$            & $v_n$     & Total         \\
Interior      & 1.287950E-03        & 1.509930E-03 & 1.362890E-03 & 1.462330E-03     & 1.363670E-03 & 1.431510E-03 \\
Extrapolation & 9.076977E-02        & 1.046573E-01 & 9.543207E-02 & 2.730460E-03     & 2.687530E-03 & 2.716800E-03 
\end{tblr}
\end{table}

\subsection{Multi-soliton solutions of self-dual network equation under the zero background}
In this section's numerical experiments, we will continue to discuss the initial-boundary value problem~(\ref{sd}). The $N$-soliton solution of Toda lattice under the zero background $I_n=V_n=0$ also can be given by Eq.~(\ref{dt_sd}) and $A_n^{(0)}=\Delta A_n^{(0)}/\Delta_n$, $B_{n}^{(0)}=\Delta B_{n}^{(0)}/\Delta_n$, where
\begin{equation}
    \Delta_n=\left|\begin{array}{cccccccc}
                     \lambda_1^{N-1} & \lambda_1^{N-2} & \cdots & 1 & \lambda_1^{N-1}\delta_{1,n} & \lambda_1^{N-2}\delta_{1,n} & \cdots & \delta_{1,n} \\
                     \lambda_2^{N-1} & \lambda_2^{N-2} & \cdots & 1 & \lambda_2^{N-1}\delta_{2,n} & \lambda_2^{N-2}\delta_{2,n} & \cdots & \delta_{2,n} \\
                     \vdots & \vdots & \vdots & \vdots & \vdots & \vdots & \vdots & \vdots \\
                     \lambda_N^{N-1} & \lambda_N^{n-2} & \cdots & 1 & \lambda_N^{n-1}\delta_{N,n} & \lambda_N^{N-2}\delta_{N,n} & \cdots & \delta_{N,n} \\
                     \lambda_1\delta_{1,n} & \lambda_1^{2}\delta_{1,n} & \cdots & \lambda_1^N\delta_{1,n} & -\lambda_1 & -\lambda_1^2 & \cdots & -\lambda_1^N  \\
                     \lambda_2\delta_{2,n} & \lambda_2^{2}\delta_{2,n} & \cdots & \lambda_2^N\delta_{2,n} & -\lambda_2 & -\lambda_2^2 & \cdots & -\lambda_2^N  \\
                     \vdots & \vdots & \vdots & \vdots & \vdots & \vdots & \vdots & \vdots \\
                     \lambda_N\delta_{N,n} & \lambda_N^{2}\delta_{N,n} & \cdots & \lambda_N^N\delta_{N,n} & -\lambda_N & -\lambda_N^2 & \cdots & -\lambda_N^N
                   \end{array}\right|, \ \delta_{i,n}=\lambda_i^{-2n}e^{-\frac{(\lambda-1)(\lambda+1)t}{\lambda_i}}.
\end{equation}
Here $\Delta A_n^{(0)}$, $\Delta B_n^{(0)}$, and $\Delta B_n^{(N-1)}$ can be obtained by replacing columns $N$, $2N$, and $N+1$ of $\Delta_n$ with the vector $(-\lambda_1^N, -\lambda_2^N,...,-\lambda_N^N, -\delta_{1,n}, -\delta_{2,n},..., \delta_{N,n})^T$. Next, we primarily apply the PG-PhyCRNet method to verify the cases when $N=1$ and $N=2$, and utilize the pseudo-lattice training method to develop the one-, two-soliton model architecture of Eq.~(\ref{sd}), demonstrating good extrapolation ability.

\subsubsection{One-soliton solution}

When $N=1$ and $\lambda_1=2$, the one-soliton solution can be given by Eq.~(\ref{dt_sd}). The spatial domain $\Omega=[-12, 13]$ is divided into $325$ segments, while the time domain $[t_0, t_1]=[-1, 1]$ is divided into $200$ segments (with a training time step of $100$ and the remaining $100$ segments used for extrapolation predictions). After training for $1000$ epochs, the loss reaches $6.717000E-07$ after $698$ seconds. Similarly, training and predictions were conducted in the time domain $t\in[-1, 0]$. The $\mathbb{L}_2$ error for component $I_n$ reaches $2.672990E-03$, while that for component $V_n$ reaches $3.255200E-03$. We further conducted extrapolation predictions for the next $100$ time steps. The $\mathbb{L}_2$ error of component $I_n$ in the time domain $t\in[0, 1]$ reaches $5.417660E-03$, and for component $V_n$ reaches $6.182490E-03$.

\begin{figure*}[!htb]
    \begin{center}
    {\scalebox{0.32}[0.32]{\includegraphics{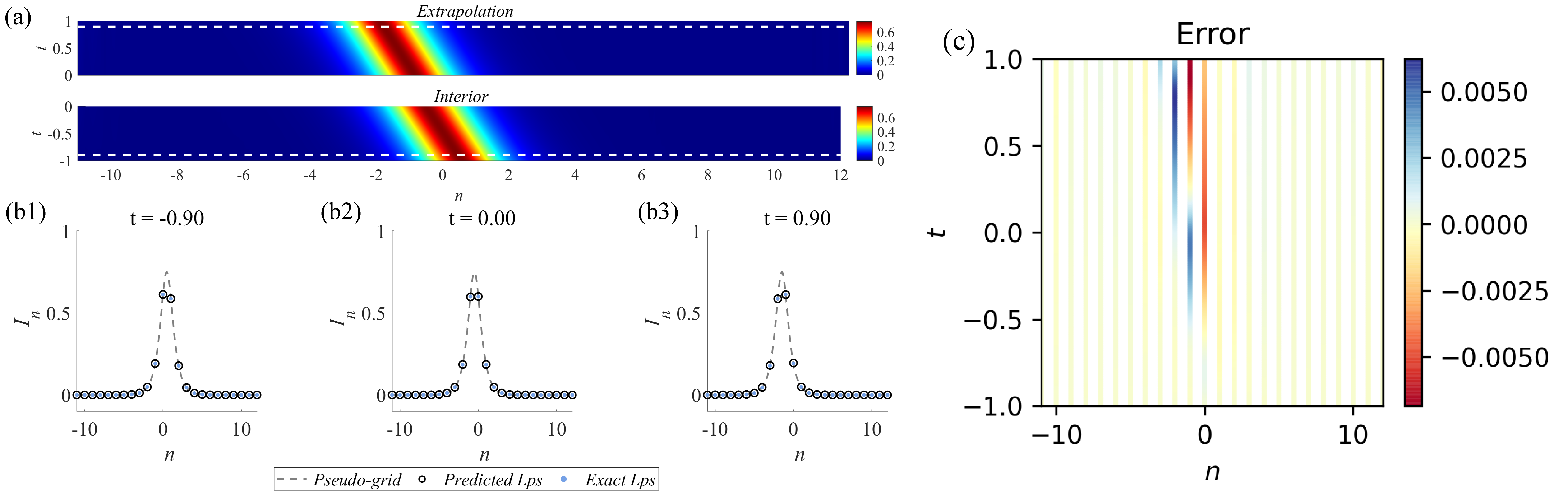}}}
    \end{center}
    \vspace{-0.15in} \caption{\small The numerical one-soliton solution of self-dual network equation for the component $I_n$ is presented. (a) illustrates the density plot of component $I_n$, with the lower sub-figure depicting the prediction within the time domain $t\in[-1, 0]$, and the upper sub-figure representing the extrapolated prediction within the time domain $t\in[0, 1]$. The white dashed lines indicate the selected moments at $t = -0.9$ and $t = 0.9$; (b1)-(b3) show the temporal evolution of one-soliton solution at three different moments. (c) presents the density plot of the lattice error (predicted solution minus exact solution).} \label{sd1_u}
\end{figure*}

\begin{figure*}[!htb]
    \begin{center}
    {\scalebox{0.32}[0.32]{\includegraphics{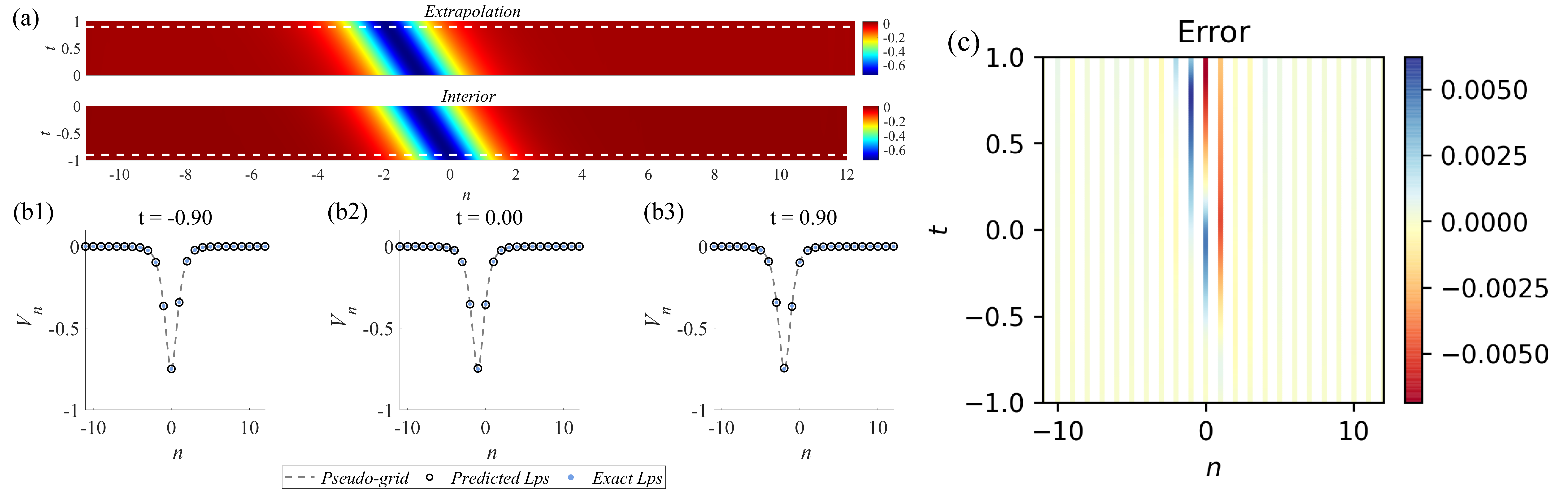}}}
    \end{center}
    \vspace{-0.15in} \caption{\small The numerical one-soliton solution of self-dual network equation for the component $V_n$ is presented. (a) illustrates the density plot of component $V_n$, with the lower sub-figure depicting the prediction within the time domain $t\in[-1, 0]$, and the upper sub-figure representing the extrapolated prediction within the time domain $t\in[0, 1]$. The white dashed lines indicate the selected moments at $t = -0.9$ and $t = 0.9$; (b1)-(b3) show the temporal evolution of one-soliton solution at three different moments. (c) presents the density plot of the lattice error (predicted solution minus exact solution).} \label{sd1_v}
\end{figure*}

\begin{figure*}[!htb]
    \begin{center}
    {\scalebox{0.32}[0.32]{\includegraphics{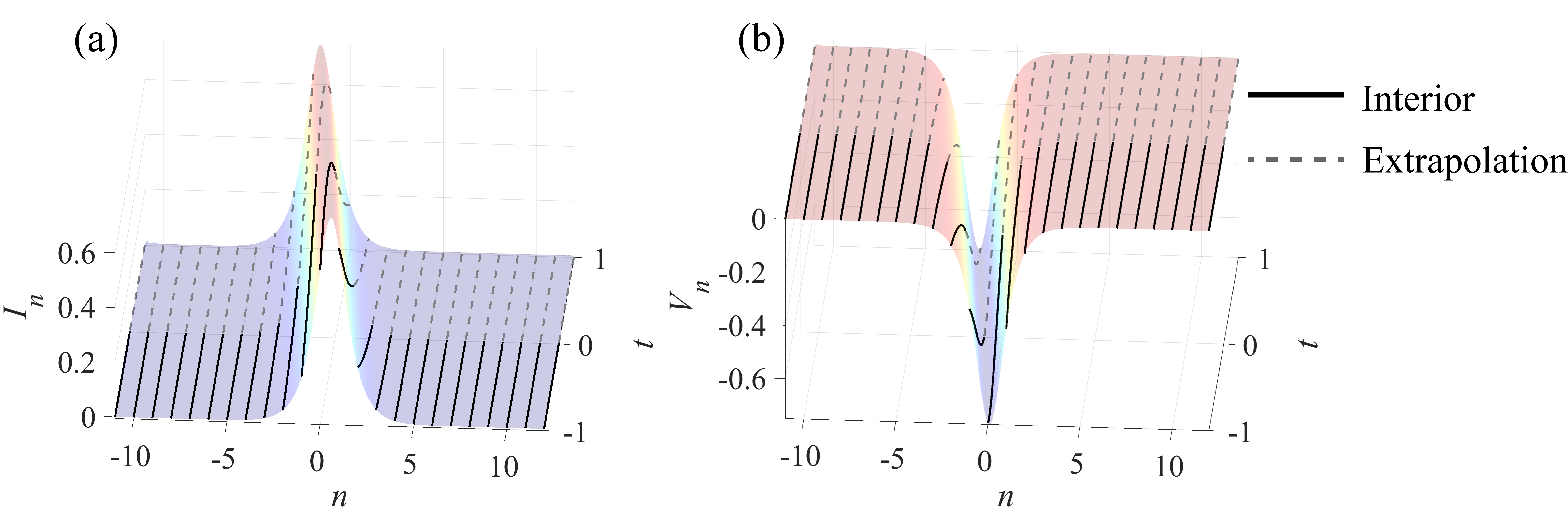}}}
    \end{center}
    \vspace{-0.15in} \caption{\small Three-dimensional plots of the numerical solution of one-soliton for self-dual network equation, where the solid black line indicates the prediction within the time domain $t\in[-1, 0]$, and the grey dashed line indicates the extrapolated prediction within the time domain $t\in[0, 1]$. Colored surfaces represent predicted solutions with pseudo grid.} \label{sd1_3d}
\end{figure*}

Figs. \ref{sd1_u} and \ref{sd1_v} present the density plots, temporal evolution plots, and error dynamics plots of the one-soliton solutions for Eq.~(\ref{sd}) based on the PG-PhyCRNet method with pseudo grid. In (a), a line soliton can be seen moving in the negative spatial direction over time. Figures (b1)-(b3) depict the temporal evolution of lattice points with the aid of pseudo grid (grey dashed lines); (b1) and (b2) show predictions within the training region, while (b3) shows extrapolation results outside the time domain. The close match between predicted and exact solutions demonstrates that the model performs well in learning the one-soliton solution. Additionally, Fig. \ref{sd1_3d} provides three-dimensional plots of the numerical solutions $I_n$ and $V_n$ . With an overall average generalization error of $1.431510E-03$ for predictions within the time domain $t\in[-1, 0]$ and $2.716800E-03$ for extrapolation outside the training range $t\in[0, 1]$, the neural network successfully captures the dynamic behavior of the one-soliton solution. The prediction accuracy for extrapolation is comparable to that within the region and does not exhibit significant differences.

Table.~\ref{tab2} compares the PG-PhyCRNet model's performance in learning the one-soliton solution of Eq.~(\ref{sd}) with and without the assistance of pseudo grid. Similar to the numerical results in the previous subsection, the PG-PhyCRNet method assisted by pseudo grid performs significantly better in capturing the lattice point features outside the time domain when training the one-soliton of Eq.~(\ref{sd}). It is observed that the one-soliton with a narrower wave width is more effectively learned using the PG-PhyCRNet method with pseudo grid assistance. In contrast, the one-soliton solution trained using PhyCRNet without pseudo grid assistance is poorly predicted outside the time domain.
\begin{table}
\centering
\caption{Errors of one-soliton solution of self-dual network equation with or without the assistance of pseudo grid training (1000 epoch)}
\label{tab2}
\begin{tblr}{
  cell{1}{2} = {c=3}{c},
  cell{1}{5} = {c=3}{c},
  hline{1,5} = {-}{0.1em},
  hline{2} = {2-7}{},
  hline{3} = {-}{},
}
              & Without~pseudo grid (681s for training time)&              &              & With~pseudo grid (698s for training time)&              &              \\
              & $u_n$            & $v_n$     & Total        & $u_n$            & $v_n$     & Total         \\
Interior      & 2.973470E-03        & 2.349200E-03 & 2.679000E-03 & 2.672990E-03     & 3.255200E-03 & 2.978890E-03 \\
Extrapolation & 2.519825E-01       & 2.248192E-01 & 2.387621E-01 & 5.417660E-03     & 6.182490E-03 & 5.813380E-03
\end{tblr}
\end{table}

\subsubsection{Two-soliton solution}
When $N=2$, $\lambda_1=1/5$ and $\lambda_2=6$, the two-soliton solution also can be given by Eq.~(\ref{dt_sd}). The spatial domain $\Omega=[-12, 13]$ is divided into $425$ segments, while the time domain $[t_0, t_1]=[-0.5, 0.5]$ is divided into $200$ segments (with a training time step of $100$ and the remaining $100$ segments used for extrapolation predictions). After training for $1000$ epochs, the loss reaches $1.678600E-06$ after $702$ seconds. Similarly, training and predictions were conducted in the time domain $t\in[-0.5, 0]$. The $\mathbb{L}_2$ error for component $I_n$ reaches $1.162890E-03$, while that for component $V_n$ reaches $7.419600E-04$. We further conducted extrapolation predictions for the next $100$ time steps. The $\mathbb{L}_2$ error of component $I_n$ in the time domain $t\in[0, 0.5]$ reaches $9.618850E-03$, and for component $V_n$ reaches $1.556281E-02$.

\begin{figure*}[!htb]
    \begin{center}
    {\scalebox{0.32}[0.32]{\includegraphics{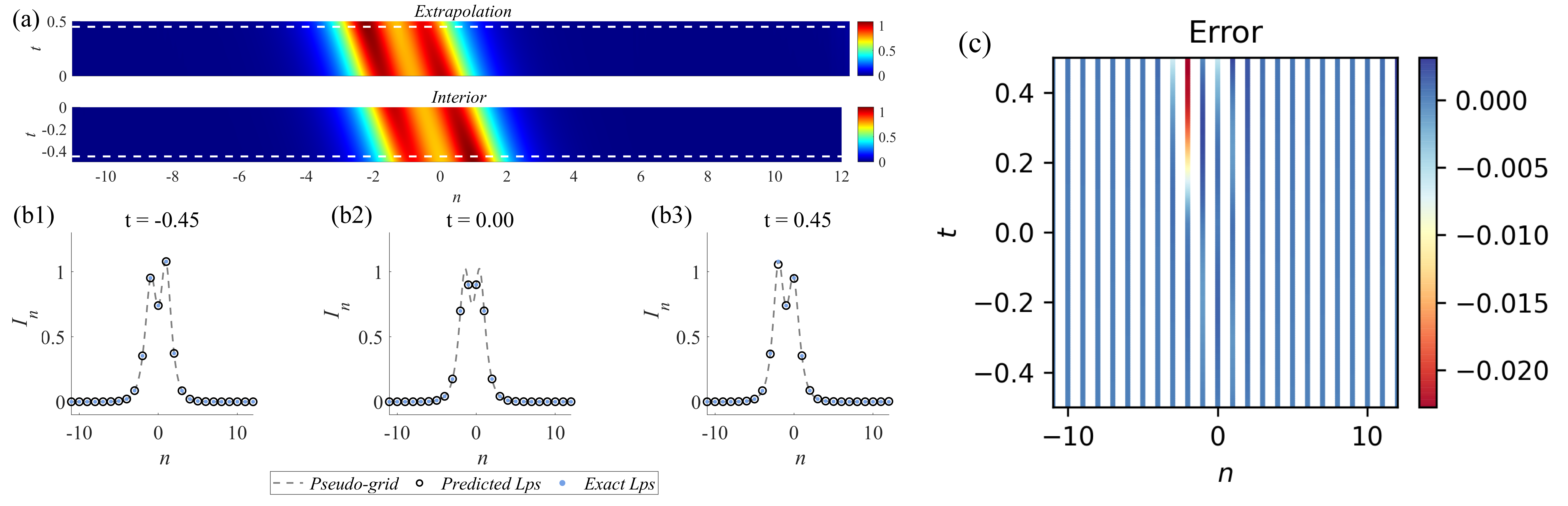}}}
    \end{center}
    \vspace{-0.15in} \caption{\small The numerical two-soliton solution of self-dual network equation for the component $I_n$ is presented. (a) illustrates the density plot of component $I_n$, with the lower sub-figure depicting the prediction within the time domain $t\in[-0.5, 0]$, and the upper sub-figure representing the extrapolated prediction within the time domain $t\in[0, 0.5]$. The white dashed lines indicate the selected moments at $t = -0.45$ and $t = 0.45$; (b1)-(b3) show the temporal evolution of two-soliton solution at three different moments. (c) presents the density plot of the lattice error (predicted solution minus exact solution).} \label{sd2_u}
\end{figure*}

\begin{figure*}[!htb]
    \begin{center}
    {\scalebox{0.32}[0.32]{\includegraphics{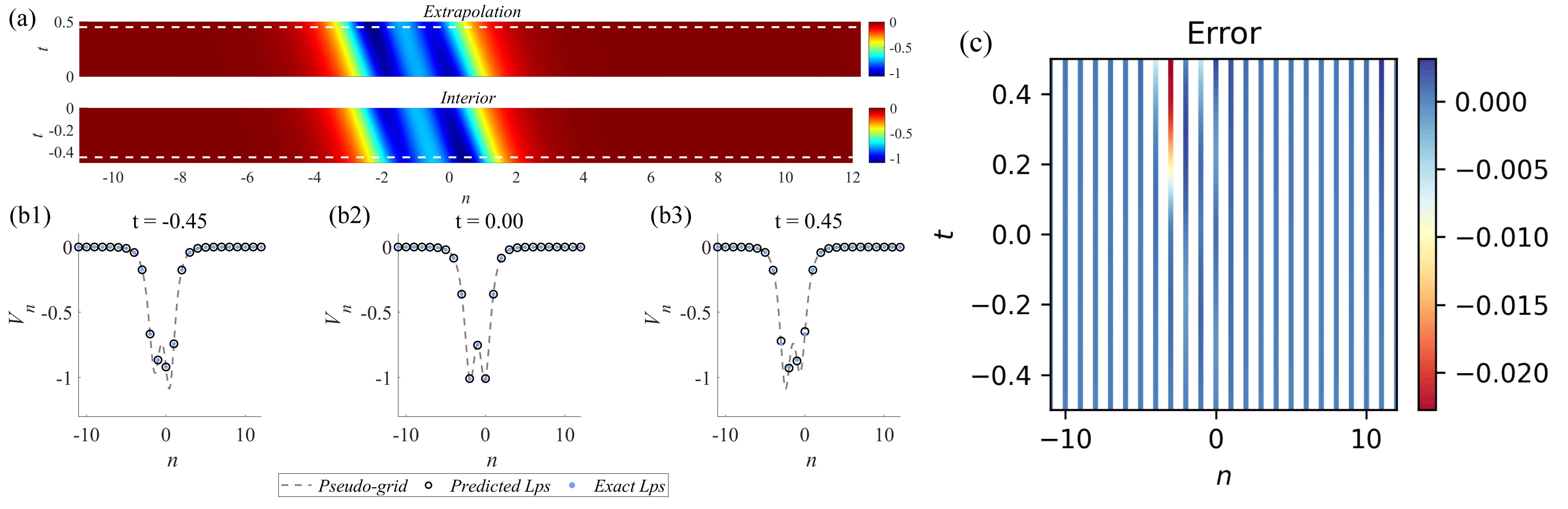}}}
    \end{center}
    \vspace{-0.15in} \caption{\small The numerical two-soliton solution of self-dual network equation for the component $V_n$ is presented. (a) illustrates the density plot of component $V_n$, with the lower sub-figure depicting the prediction within the time domain $t\in[-0.5, 0]$, and the upper sub-figure representing the extrapolated prediction within the time domain $t\in[0, 0.5]$. The white dashed lines indicate the selected moments at $t = -0.45$ and $t = 0.45$; (b1)-(b3) show the temporal evolution of two-soliton solution at three different moments. (c) presents the density plot of the lattice error (predicted solution minus exact solution).} \label{sd2_v}
\end{figure*}

\begin{figure*}[!htb]
    \begin{center}
    {\scalebox{0.32}[0.32]{\includegraphics{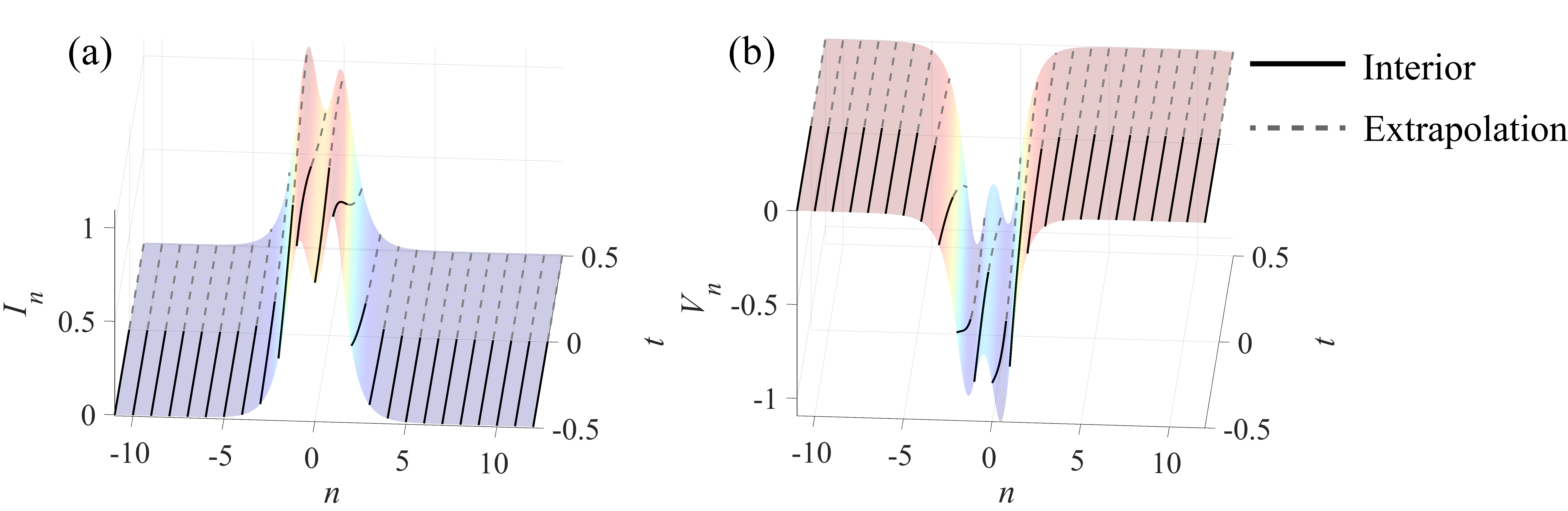}}}
    \end{center}
    \vspace{-0.15in} \caption{\small Three-dimensional plots of the numerical solution of two-soliton for self-dual network equation, where the solid black line indicates the prediction within the time domain $t\in[-0.5, 0]$, and the grey dashed line indicates the extrapolated prediction within the time domain $t\in[0, 0.5]$. Colored surfaces represent predicted solutions with pseudo grid} \label{sd2_3d}
\end{figure*}

Figs. \ref{sd2_u} and \ref{sd2_v} present the density plots, temporal evolution plots, and error dynamics plots of the two-soliton solutions for Eq.~(\ref{sd}) based on the PG-PhyCRNet method with pseudo grid. In (a), two soliton can be seen moving in the negative spatial direction over time. (b1)-(b3) depict the temporal evolution of lattice points with the aid of pseudo grid (grey dashed lines); (b1) and (b2) show predictions within the training region, while (b3) shows extrapolation results outside the time domain. The close match between predicted and exact solutions demonstrates that the model performs well in learning the two-soliton solution. Additionally, Fig. \ref{sd2_3d} provides three-dimensional plots of the numerical solutions $I_n$ and $V_n$ . With an overall average generalization error of $9.763600E-04$ for predictions within the time domain $t\in[-0.5, 0]$ and $1.292342E-02$ for extrapolation outside the training range $t\in[0, 0.5]$, the neural network successfully captures the dynamic behavior of the two-soliton solution. The prediction accuracy for extrapolation is comparable to that within the region and does not exhibit significant differences.

Table.~\ref{tab3} compares the PG-PhyCRNet model's performance in learning the two-soliton solution of Eq.~(\ref{sd}) with and without the assistance of pseudo grid. Similar to the numerical results in the previous subsection, the PG-PhyCRNet method assisted by pseudo grid performs significantly better in capturing the lattice point features outside the time domain when training the two-soliton of Eq.~(\ref{sd}). It is observed that the one-soliton with a narrower wave width is more effectively learned using the PG-PhyCRNet method with pseudo grid assistance. In contrast, the two-soliton solution trained using PhyCRNet without pseudo grid assistance is poorly predicted outside the time domain.
\begin{table}
\centering
\caption{Errors of two-soliton solution of self-dual network equation with or without the assistance of pseudo grid training (1000 epoch)}
\label{tab3}
\begin{tblr}{
  cell{1}{2} = {c=3}{c},
  cell{1}{5} = {c=3}{c},
  hline{1,5} = {-}{0.1em},
  hline{2} = {2-7}{},
  hline{3} = {-}{},
}
              & Without~pseudo grid (674s for training time)&              &              & With~pseudo grid (702s for training time)&              &              \\
              & $u_n$            & $v_n$     & Total        & $u_n$            & $v_n$     & Total         \\
Interior      & 6.788400E-04        & 4.168900E-04 & 5.639000E-04 & 1.162890E-03     & 7.419600E-04 & 9.763600E-04 \\
Extrapolation & 6.503445E-02       & 1.116403E-01 & 9.125480E-02 & 9.618850E-03     & 1.556281E-02 & 1.292342E-02
\end{tblr}
\end{table}

\section{Analysis and discussion}\label{sec5}

\subsection{The effect of random initialization and spatial domain griding}
Since the weights of the neural network are initialized using Xavier initialization, the setting of the parameter seed in the code will affect the numerical results. Besides, the integer lattice points learn with boundary information, whereas the pseudo grid lack information during training, resulting in a high degree of freedom in their numerical solutions. To investigate the stability of integer lattice points in the presence of the pseudo grid in PG-PhyCRNet, numerical experiments with different randomly generated seed values and numbers of pseudo grid were conducted to observe their impact on the results (repeated experiments are shown in Tables.~\ref{tab5}-\ref{tab19} in the appendix):

\noindent (1) The $\mathbb{L}_2$ generalization errors obtained under five different random seeds do not differ significantly, indicating that the degree of freedom of pseudo grid does not greatly impact the prediction and extrapolation of integer lattice points in multiple repeated experiments. Although some accuracy within the time domain may be lost or oscillations may appear in the background waves on pseudo grid, the PG-PhyCRNet method with the aid of pseudo grid still performs well in predicting solitary waves as mentioned in this paper.

\begin{figure*}[!htb]
    \begin{center}
    {\scalebox{0.4}[0.4]{\includegraphics{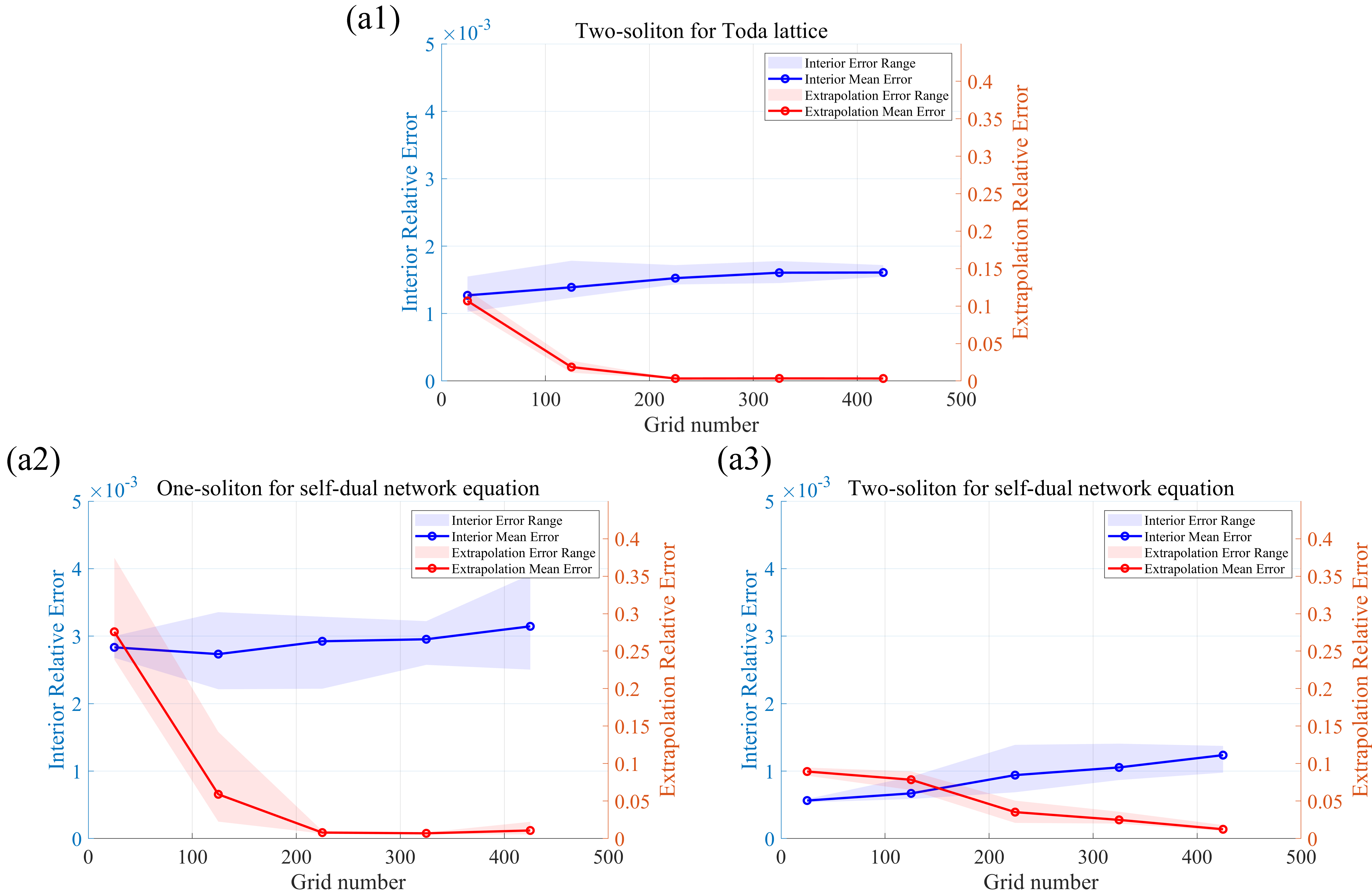}}}
    \end{center}
    \vspace{-0.15in} \caption{\small Effect of spatial griding on the prediction of soliton solutions of the Toda lattice and self-dual network equations under different seeds. The red region indicates the $\mathbb{L}_2$ error range predicted inside the given time domain, while the blue region indicates the $\mathbb{L}_2$ error range predicted outside the time domain.} \label{grid}
\end{figure*}

\noindent (2) The prediction of different soliton solutions is affected by the density of pseudo grid. In Fig. \ref{grid}, we show the effect of spatial griding on the prediction of soliton solutions of the Toda lattice and self-dual network equations under different seeds. It can be observed that as the density of pseudo grid increases, the prediction error within the given spatiotemporal region increases until it stabilizes. Conversely, the extrapolation prediction error outside the time region decreases until it stabilizes. Therefore, it is necessary to choose an appropriate number of pseudo grid to participate in the model training.

\subsection{Remark}
The presence of excessive pseudo grid can lead to their weights being overly emphasized in the loss function, while the weights of integer lattice points are relatively diminished. Consequently, there is a certain loss when learning the integer lattice points within the spatiotemporal domain. One significant characteristic of CNNs is weight sharing, which imposes constraints on the weights of convolutional kernels by the integer lattice points during the training process. Therefore, the accuracy of learning the integer lattice points within the spatiotemporal domain does not suffer severely. However, the existence of pseudo grid makes it easier for the convolutional kernels to capture the dynamic behavioral characteristics of solitons. In other words, the PG-PhyCRNet method trained with pseudo grid learns a regression problem of curves rather than the dynamics of individual lattice points. Consequently, when the width of a soliton is small, too few trend characteristics of integer lattice points are captured at its steep position (where the slope of the function at a certain point is sufficiently large). This explains why the extrapolated performance of PG-PhyCRNet without the assistance of pseudo grid is poor or significantly worse (see Tables.~\ref{tab5}-\ref{tab19}). Conversely, PG-PhyCRNet with the assistance of pseudo grid makes it easier for the convolutional kernels to capture the evolutionary trend of solitons, thereby enhancing the extrapolation capability of integer lattice points beyond the time domain.

\section{Conclusions}
In conventional discrete learning methods, most discussions focus on continuous nonlinear PDEs. The dynamics of the solutions to the corresponding continuous equations are indirectly explored by directly discretizing (using difference schemes) the continuous equations. Therefore, another discrete format of nonlinear integrable equations: integrable NLEs, are discussed via the discrete deep learning methods (PhyCRNet method) in this paper. These NLEs avoid the stability and convergence issues that arise from discretizing continuous equations with finite difference schemes in numerical calculations.

In integrable NLEs, the translation operator replaces the higher-order dispersion terms in continuous equations. Therefore, based on the ideas presented in Ref.~\cite{phycrnet}, we propose the PG-PhyCRNet method specifically designed for solving integrable lattice equations. This method also cleverly addresses the difficulties that different waveforms of lattice solitons present to discrete learning training. The addition of pseudo grid enables the model to comprehensively learn various lattice soliton solutions:
\begin{itemize}
  \item For lattice solitons with slower wave speeds and smoother waveforms, we train and predict directly on integer lattice points. In numerical experiments, we used a one-soliton of the Toda lattice and rational solitons of the self-dual network equation. The numerical results show that both the internal predictions and extrapolated predictions are excellent within the given space-time domain.
  \item For lattice solitons with faster wave speeds and steeper waveforms, we train and predict using both pseudo grid and integer lattice points. In numerical experiments, we used two-solitons of the Toda lattice and one- and two-solitons of the self-dual network equation. The numerical results indicate that the method of training with pseudo grid significantly outperforms the method without pseudo grid in terms of extrapolation capability.
\end{itemize}

The weight sharing in CNNs alleviates the errors' balance challenge between pseudo grid and integer lattice points in PG-PhyCRNet training. Through the repeated experiments, PG-PhyCRNet with pseudo grid method sacrifices a small portion of training errors in the spatiotemporal domain compared to the version without pseudo grid. However, it exhibits excellent performance in capturing soliton trends and enhancing the model's extrapolation capability for solitons. 

The PG-PhyCRNet method proposed for integrable NLEs is very necessary, as it provides a new perspective for solving continuous integrable systems using discrete deep learning methods. Of course, the numerical experiments in this paper only involve solving different types and structures of solitons. It is worth studying the generalization capability of this model for more complex localized wave solutions, such as rogue waves, breathers, and interaction solutions. Additionally, embedding certain properties of integrable discrete equations (such as discrete conservation laws, discrete Hamiltonian structures, continuous limits, and recursion operators) into the network to achieve different goals is something we aim to explore in the future. 

\section*{CRediT authorship contribution statement} 
{\bf Zhe Lin}: Writing - original draft, Writing - review $\&$ editing, Methodology, Software. {\bf Yong Chen}: Supervision, Writing - original draft, Writing - review $\&$ editing, Methodology, Project administration.

\section*{Declaration of competing interest}
The authors declare that they have no known competing financial interests or personal relationships that could have appeared to influence the work reported in this paper.

\section*{Acknowledgments} 
The project is supported by the National Natural Science Foundation of China (No. 12175069 and No. 12235007), Science and Technology Commission of Shanghai Municipality, China (No. 21JC1402500 and No. 22DZ2229014), and Natural Science Foundation of Shanghai, China (No. 23ZR1418100).

\section*{Data availability} 
The data that support the findings of this study are available from the corresponding author upon reasonable request.

\bibliography{apssamp}

\appendix
\section{Repeated experimental and relative $\mathbb{L}_2$ errors for training two-soliton solution for Toda lattice in the spatial domain $[-12,13]$}
\begin{table}[H]
\centering
\caption{Based on $25$ grid numbers' training }
\label{tab5}
\begin{tblr}{
  cell{1}{1} = {c=1,r=2}{},
  cell{1}{2} = {c=3}{c},
  cell{1}{5} = {c=3}{c},
  hline{1,8} = {-}{0.1em},
  hline{2} = {2-7}{},
  hline{3} = {-}{},
}
     Seed         & Interior &              &              & Extrapolation&              &              \\
              & $u_n$            & $v_n$     & Total        & $u_n$            & $v_n$     & Total         \\
100 & 1.004390E-03 & 1.066160E-03 & 1.024560E-03 & 9.199270E-02 & 1.062825E-01 & 9.679334E-02 \\
1100 & 1.141840E-03 & 1.187920E-03 & 1.156780E-03 & 9.432476E-02 & 1.106409E-01 & 9.983428E-02 \\
2100 & 1.287950E-03 & 1.509930E-03 & 1.362890E-03 & 9.076977E-02 & 1.046573E-01 & 9.543207E-02 \\ 
3100 & 1.166190E-03 & 1.433770E-03 & 1.257990E-03 & 1.198982E-01 & 1.246799E-01 & 1.214482E-01 \\ 
4100 & 1.620150E-03 & 1.387780E-03 & 1.549620E-03 & 1.182343E-01 & 1.264537E-01 & 1.209242E-01 \\

\end{tblr}
\end{table}
\begin{table}[H]
\centering
\caption{Based on $125$ grid numbers' training }
\label{tab6}
\begin{tblr}{
  cell{1}{1} = {c=1,r=2}{},
  cell{1}{2} = {c=3}{c},
  cell{1}{5} = {c=3}{c},
  hline{1,8} = {-}{0.1em},
  hline{2} = {2-7}{},
  hline{3} = {-}{},
}
     Seed         & Interior &              &              & Extrapolation&              &              \\
              & $u_n$            & $v_n$     & Total        & $u_n$            & $v_n$     & Total         \\
100 & 1.358250E-03 & 1.373920E-03 & 1.363290E-03 & 9.823230E-03 & 1.318727E-02 & 1.101165E-02\\
1100 & 1.290880E-03 & 1.225100E-03 & 1.270210E-03 & 2.419791E-02 & 3.235194E-02 & 2.707461E-02\\
2100 & 1.295460E-03 & 1.089890E-03 & 1.233440E-03 & 1.211099E-02 & 1.750363E-02 & 1.406270E-02\\
3100 & 1.346560E-03 & 1.188820E-03 & 1.298190E-03 & 1.712007E-02 & 2.405495E-02 & 1.960698E-02\\
4100 & 1.891910E-03 & 1.528060E-03 & 1.783620E-03 & 1.975810E-02 & 2.388229E-02 & 2.116490E-02\\

\end{tblr}
\end{table}
\begin{table}[H]
\centering
\caption{Based on $225$ grid numbers' training }
\label{tab7}
\begin{tblr}{
  cell{1}{1} = {c=1,r=2}{},
  cell{1}{2} = {c=3}{c},
  cell{1}{5} = {c=3}{c},
  hline{1,8} = {-}{0.1em},
  hline{2} = {2-7}{},
  hline{3} = {-}{},
}
     Seed         & Interior &              &              & Extrapolation&              &              \\
              & $u_n$            & $v_n$     & Total        & $u_n$            & $v_n$     & Total         \\
100 & 1.454350E-03 & 1.575410E-03 & 1.494140E-03 & 3.424500E-03 & 3.625440E-03 & 3.490030E-03\\
1100 & 1.783300E-03 & 1.573400E-03 & 1.718950E-03 & 3.590040E-03 & 3.274400E-03 & 3.492180E-03\\
2100 & 1.462330E-03 & 1.363670E-03 & 1.431510E-03 & 2.730460E-03 & 2.687530E-03 & 2.716800E-03\\
3100 & 1.462900E-03 & 1.434220E-03 & 1.453790E-03 & 3.558970E-03 & 3.347580E-03 & 3.492750E-03\\
4100 & 1.597290E-03 & 1.370320E-03 & 1.528360E-03 & 3.498280E-03 & 3.283320E-03 & 3.430990E-03\\

\end{tblr}
\end{table}
\begin{table}[H]
\centering
\caption{Based on $325$ grid numbers' training }
\label{tab8}
\begin{tblr}{
  cell{1}{1} = {c=1,r=2}{},
  cell{1}{2} = {c=3}{c},
  cell{1}{5} = {c=3}{c},
  hline{1,8} = {-}{0.1em},
  hline{2} = {2-7}{},
  hline{3} = {-}{},
}
     Seed         & Interior &              &              & Extrapolation&              &              \\
              & $u_n$            & $v_n$     & Total        & $u_n$            & $v_n$     & Total         \\
100 & 1.789990E-03 & 1.458460E-03 & 1.691030E-03 & 3.454720E-03 & 3.451360E-03 & 3.453650E-03\\
1100 & 1.762830E-03 & 1.286240E-03 & 1.625650E-03 & 3.959900E-03 & 2.963380E-03 & 3.670690E-03\\
2100 & 1.523710E-03 & 1.285470E-03 & 1.451770E-03 & 2.942450E-03 & 3.027140E-03 & 2.969800E-03\\
3100 & 1.479490E-03 & 1.489380E-03 & 1.482660E-03 & 2.669840E-03 & 3.111110E-03 & 2.818510E-03\\
4100 & 1.748590E-03 & 1.844100E-03 & 1.779700E-03 & 4.281690E-03 & 4.658000E-03 & 4.405560E-03\\

\end{tblr}
\end{table}
\begin{table}[H]
\centering
\caption{Based on $425$ grid numbers' training }
\label{tab9}
\begin{tblr}{
  cell{1}{1} = {c=1,r=2}{},
  cell{1}{2} = {c=3}{c},
  cell{1}{5} = {c=3}{c},
  hline{1,8} = {-}{0.1em},
  hline{2} = {2-7}{},
  hline{3} = {-}{},
}
     Seed         & Interior &              &              & Extrapolation&              &              \\
              & $u_n$            & $v_n$     & Total        & $u_n$            & $v_n$     & Total         \\
100 & 1.645460E-03 & 1.670310E-03 & 1.653450E-03 & 3.941530E-03 & 3.485480E-03 & 3.801610E-03\\
1100 & 1.760540E-03 & 1.628470E-03 & 1.719400E-03 & 3.398780E-03 & 3.257960E-03 & 3.354380E-03\\
2100 & 1.559400E-03 & 1.554180E-03 & 1.557730E-03 & 3.227280E-03 & 3.319680E-03 & 3.257120E-03\\
3100 & 1.570980E-03 & 1.491500E-03 & 1.546000E-03 & 2.995290E-03 & 3.611780E-03 & 3.205410E-03\\
4100 & 1.588130E-03 & 1.524420E-03 & 1.568030E-03 & 2.973770E-03 & 3.200140E-03 & 3.048010E-03\\

\end{tblr}
\end{table}

\section{Repeated experimental and relative $\mathbb{L}_2$ errors for training one-soliton solution for self-dual network equation in the spatial domain $[-12,13]$}
\begin{table}[H]
\centering
\caption{Based on $25$ grid numbers' training }
\label{tab10}
\begin{tblr}{
  cell{1}{1} = {c=1,r=2}{},
  cell{1}{2} = {c=3}{c},
  cell{1}{5} = {c=3}{c},
  hline{1,8} = {-}{0.1em},
  hline{2} = {2-7}{},
  hline{3} = {-}{},
}
     Seed         & Interior &              &              & Extrapolation&              &              \\
              & $I_n$            & $V_n$     & Total        & $T_n$            & $V_n$     & Total         \\
100 & 2.973470E-03 & 2.349200E-03 & 2.679000E-03 & 2.519825E-01 & 2.248192E-01 & 2.387621E-01\\
1100 & 3.114740E-03 & 2.687380E-03 & 2.908520E-03 & 3.546937E-01 & 3.933042E-01 & 3.745328E-01\\
2100 & 2.545430E-03 & 3.396040E-03 & 3.001810E-03 & 2.538650E-01 & 2.427684E-01 & 2.483683E-01\\
3100 & 2.603690E-03 & 3.033760E-03 & 2.827310E-03 & 2.831439E-01 & 2.339709E-01 & 2.596781E-01\\
4100 & 2.919450E-03 & 2.578170E-03 & 2.753790E-03 & 2.595958E-01 & 2.557138E-01 & 2.576585E-01\\

\end{tblr}
\end{table}
\begin{table}[H]
\centering
\caption{Based on $125$ grid numbers' training }
\label{tab11}
\begin{tblr}{
  cell{1}{1} = {c=1,r=2}{},
  cell{1}{2} = {c=3}{c},
  cell{1}{5} = {c=3}{c},
  hline{1,8} = {-}{0.1em},
  hline{2} = {2-7}{},
  hline{3} = {-}{},
}
     Seed         & Interior &              &              & Extrapolation&              &              \\
              & $I_n$            & $V_n$     & Total        & $I_n$            & $V_n$     & Total         \\
100 & 2.089710E-03 & 3.179500E-03 & 2.691360E-03 & 2.655297E-02 & 1.768221E-02 & 2.254981E-02\\
1100 & 2.173330E-03 & 2.301320E-03 & 2.238360E-03 & 6.974361E-02 & 3.462078E-02 & 5.502694E-02\\
2100 & 2.144920E-03 & 2.278520E-03 & 2.212860E-03 & 3.404159E-02 & 4.308837E-02 & 3.883774E-02\\
3100 & 2.269970E-03 & 3.865790E-03 & 3.171380E-03 & 1.751873E-01 & 9.917303E-02 & 1.422795E-01\\
4100 & 3.288780E-03 & 3.420710E-03 & 3.355510E-03 & 3.574693E-02 & 3.701973E-02 & 3.639008E-02\\

\end{tblr}
\end{table}
\begin{table}[H]
\centering
\caption{Based on $225$ grid numbers' training }
\label{tab12}
\begin{tblr}{
  cell{1}{1} = {c=1,r=2}{},
  cell{1}{2} = {c=3}{c},
  cell{1}{5} = {c=3}{c},
  hline{1,8} = {-}{0.1em},
  hline{2} = {2-7}{},
  hline{3} = {-}{},
}
     Seed         & Interior &              &              & Extrapolation&              &              \\
              & $I_n$            & $V_n$     & Total        & $T_n$            & $V_n$     & Total          \\
100 & 2.028130E-03 & 3.893660E-03 & 3.106000E-03 & 9.274210E-03 & 8.104070E-03 & 8.707720E-03\\
1100 & 1.755860E-03 & 2.602480E-03 & 2.220680E-03 & 5.947070E-03 & 7.775850E-03 & 6.923810E-03\\
2100 & 2.720080E-03 & 3.534910E-03 & 3.154680E-03 & 6.701000E-03 & 7.475810E-03 & 7.099710E-03\\
3100 & 2.570200E-03 & 3.872700E-03 & 3.287810E-03 & 1.070728E-02 & 9.861580E-03 & 1.029233E-02\\
4100 & 2.356570E-03 & 3.270510E-03 & 2.851250E-03 & 6.671430E-03 & 6.627350E-03 & 6.649380E-03\\

\end{tblr}
\end{table}
\begin{table}[H]
\centering
\caption{Based on $325$ grid numbers' training }
\label{tab13}
\begin{tblr}{
  cell{1}{1} = {c=1,r=2}{},
  cell{1}{2} = {c=3}{c},
  cell{1}{5} = {c=3}{c},
  hline{1,8} = {-}{0.1em},
  hline{2} = {2-7}{},
  hline{3} = {-}{},
}
     Seed         & Interior &              &              & Extrapolation&              &              \\
              & $I_n$            & $V_n$     & Total        & $T_n$            & $V_n$     & Total          \\
100 & 2.672990E-03 & 3.255200E-03 & 2.978890E-03 & 5.417660E-03 & 6.182490E-03 & 5.813380E-03\\
1100 & 2.551020E-03 & 3.415600E-03 & 3.015270E-03 & 6.496750E-03 & 8.152780E-03 & 7.372960E-03\\
2100 & 2.662000E-03 & 3.262530E-03 & 2.978000E-03 & 5.536020E-03 & 6.573370E-03 & 6.077840E-03\\
3100 & 2.680300E-03 & 3.684550E-03 & 3.222720E-03 & 8.051520E-03 & 8.241780E-03 & 8.147390E-03\\
4100 & 1.864270E-03 & 3.125680E-03 & 2.574600E-03 & 6.860780E-03 & 8.198070E-03 & 7.560300E-03\\

\end{tblr}
\end{table}
\begin{table}[H]
\centering
\caption{Based on $425$ grid numbers' training }
\label{tab14}
\begin{tblr}{
  cell{1}{1} = {c=1,r=2}{},
  cell{1}{2} = {c=3}{c},
  cell{1}{5} = {c=3}{c},
  hline{1,8} = {-}{0.1em},
  hline{2} = {2-7}{},
  hline{3} = {-}{},
}
     Seed         & Interior &              &              & Extrapolation&              &              \\
              & $I_n$            & $V_n$     & Total        & $T_n$            & $V_n$     & Total          \\
100 & 4.203370E-03 & 3.599190E-03 & 3.912400E-03 & 1.030187E-02 & 9.803130E-03 & 1.005513E-02\\
1100 & 2.693260E-03 & 3.412580E-03 & 3.074700E-03 & 6.672860E-03 & 6.490600E-03 & 6.582190E-03\\
2100 & 2.395680E-03 & 3.745470E-03 & 3.145100E-03 & 1.060885E-02 & 8.064490E-03 & 9.420590E-03\\
3100 & 2.415120E-03 & 3.642350E-03 & 3.091390E-03 & 2.550555E-02 & 1.864582E-02 & 2.233421E-02\\
4100 & 2.034670E-03 & 2.896380E-03 & 2.503680E-03 & 5.263380E-03 & 5.644690E-03 & 5.457720E-03\\

\end{tblr}
\end{table}

\section{Repeated experimental and relative $\mathbb{L}_2$ errors for training two-soliton solution for self-dual network equation in the spatial domain $[-12,13]$}
\begin{table}[H]
\centering
\caption{Based on $25$ grid numbers' training }
\label{tab15}
\begin{tblr}{
  cell{1}{1} = {c=1,r=2}{},
  cell{1}{2} = {c=3}{c},
  cell{1}{5} = {c=3}{c},
  hline{1,8} = {-}{0.1em},
  hline{2} = {2-7}{},
  hline{3} = {-}{},
}
     Seed         & Interior &              &              & Extrapolation&              &              \\
              & $I_n$            & $V_n$     & Total        & $T_n$            & $V_n$     & Total          \\
100 & 6.723700E-04 & 3.469900E-04 & 5.357300E-04 & 6.196126E-02 & 1.012896E-01 & 8.387203E-02\\
1100 & 6.788400E-04 & 4.168900E-04 & 5.639000E-04 & 6.503445E-02 & 1.116403E-01 & 9.125480E-02\\
2100 & 6.811500E-04 & 4.732600E-04 & 5.869600E-04 & 6.264318E-02 & 1.059869E-01 & 8.695842E-02\\
3100 & 6.901300E-04 & 3.579200E-04 & 5.504500E-04 & 6.329838E-02 & 1.111489E-01 & 9.033848E-02\\
4100 & 6.852900E-04 & 4.495200E-04 & 5.800600E-04 & 6.519826E-02 & 1.171786E-01 & 9.470403E-02\\

\end{tblr}
\end{table}
\begin{table}[H]
\centering
\caption{Based on $125$ grid numbers' training }
\label{tab16}
\begin{tblr}{
  cell{1}{1} = {c=1,r=2}{},
  cell{1}{2} = {c=3}{c},
  cell{1}{5} = {c=3}{c},
  hline{1,8} = {-}{0.1em},
  hline{2} = {2-7}{},
  hline{3} = {-}{},
}
     Seed         & Interior &              &              & Extrapolation&              &              \\
              & $I_n$            & $V_n$     & Total        & $T_n$            & $V_n$     & Total          \\
100 & 6.849500E-04 & 5.057600E-04 & 6.024700E-04 & 5.959053E-02 & 1.000647E-01 & 8.226188E-02\\
1100 & 6.144300E-04 & 6.327200E-04 & 6.236000E-04 & 4.975587E-02 & 1.059173E-01 & 8.262451E-02\\
2100 & 7.169700E-04 & 1.080460E-03 & 9.160800E-04 & 3.486782E-02 & 9.738489E-02 & 7.301121E-02\\
3100 & 7.115200E-04 & 4.221800E-04 & 5.856700E-04 & 3.314541E-02 & 8.553515E-02 & 6.475365E-02\\
4100 & 6.964000E-04 & 5.303800E-04 & 6.193600E-04 & 6.024640E-02 & 1.116978E-01 & 8.962430E-02\\

\end{tblr}
\end{table}
\begin{table}[H]
\centering
\caption{Based on $225$ grid numbers' training }
\label{tab17}
\begin{tblr}{
  cell{1}{1} = {c=1,r=2}{},
  cell{1}{2} = {c=3}{c},
  cell{1}{5} = {c=3}{c},
  hline{1,8} = {-}{0.1em},
  hline{2} = {2-7}{},
  hline{3} = {-}{},
}
     Seed         & Interior &              &              & Extrapolation&              &              \\
              & $I_n$            & $V_n$     & Total        & $T_n$            & $V_n$     & Total          \\
100 & 9.330600E-04 & 6.861400E-04 & 8.195200E-04 & 4.437688E-02 & 4.155745E-02 & 4.299681E-02\\
1100 & 6.980900E-04 & 6.734000E-04 & 6.859100E-04 & 1.297905E-02 & 2.655921E-02 & 2.087294E-02\\
2100 & 9.907600E-04 & 1.158130E-03 & 1.077310E-03 & 1.468222E-02 & 3.614822E-02 & 2.754275E-02\\
3100 & 1.015390E-03 & 1.683930E-03 & 1.388930E-03 & 4.295351E-02 & 2.244343E-02 & 3.431418E-02\\
4100 & 7.889300E-04 & 6.640200E-04 & 7.294500E-04 & 4.236627E-02 & 5.771308E-02 & 5.058945E-02\\

\end{tblr}
\end{table}
\begin{table}[H]
\centering
\caption{Based on $325$ grid numbers' training }
\label{tab18}
\begin{tblr}{
  cell{1}{1} = {c=1,r=2}{},
  cell{1}{2} = {c=3}{c},
  cell{1}{5} = {c=3}{c},
  hline{1,8} = {-}{0.1em},
  hline{2} = {2-7}{},
  hline{3} = {-}{},
}
     Seed         & Interior &              &              & Extrapolation&              &              \\
              & $I_n$            & $V_n$     & Total        & $T_n$            & $V_n$     & Total         \\
100 & 8.840300E-04 & 1.253540E-03 & 1.083790E-03 & 1.548189E-02 & 2.359871E-02 & 1.993888E-02\\
1100 & 1.013320E-03 & 1.068150E-03 & 1.040970E-03 & 1.521999E-02 & 2.901112E-02 & 2.313511E-02\\
2100 & 6.687600E-04 & 1.028990E-03 & 8.669600E-04 & 1.875190E-02 & 2.473241E-02 & 2.193309E-02\\
3100 & 1.055200E-03 & 1.688590E-03 & 1.406540E-03 & 1.844988E-02 & 2.791708E-02 & 2.364030E-02\\
4100 & 8.523800E-04 & 8.927600E-04 & 8.727100E-04 & 2.094418E-02 & 4.615326E-02 & 3.578369E-02\\

\end{tblr}
\end{table}
\begin{table}[H]
\centering
\caption{Based on $425$ grid numbers' training }
\label{tab19}
\begin{tblr}{
  cell{1}{1} = {c=1,r=2}{},
  cell{1}{2} = {c=3}{c},
  cell{1}{5} = {c=3}{c},
  hline{1,8} = {-}{0.1em},
  hline{2} = {2-7}{},
  hline{3} = {-}{},
}
     Seed         & Interior &              &              & Extrapolation&              &              \\
              & $I_n$            & $V_n$     & Total        & $T_n$            & $V_n$     & Total         \\
100 & 9.886800E-04 & 1.558600E-03 & 1.303840E-03 & 9.613700E-03 & 1.143157E-02 & 1.055762E-02\\
1100 & 1.162890E-03 & 7.419600E-04 & 9.763600E-04 & 9.618850E-03 & 1.556281E-02 & 1.292342E-02\\
2100 & 6.929300E-04 & 1.483680E-03 & 1.156170E-03 & 1.360377E-02 & 2.164958E-02 & 1.806173E-02\\
3100 & 1.106100E-03 & 1.597920E-03 & 1.373070E-03 & 9.200870E-03 & 1.103213E-02 & 1.015362E-02\\
4100 & 9.166300E-04 & 1.315060E-03 & 1.132580E-03 & 2.262608E-02 & 3.000274E-02 & 2.655471E-02\\

\end{tblr}
\end{table}
\end{document}